\documentclass[12pt,a4paper]{amsart}
\usepackage[utf8]{inputenc}
\usepackage{amsmath}
\usepackage{amssymb}
\usepackage{geometry}
\usepackage[english]{babel}
\usepackage{hyperref}
\usepackage{xcolor}
\usepackage{graphbox}
\usepackage{setspace}
\usepackage{graphicx}
\usepackage{subcaption}
\usepackage{multirow}
\usepackage{floatrow}
\newfloatcommand{capbtabbox}{table}[][\FBwidth]
\usepackage{t1enc}
\date{}

\begin{document}

\title{Field equation of thermodynamic gravity and galactic rotational curves} 
\author{M. Pszota$^{1,2,3,4}$ and P. V\'an$^{2,3,4}$ }

\address{$^1$ Eötvös Loránd University, Budapest\\
		$^2$Department of Theoretical Physics, Wigner Research Centre for Physics, H-1525 Budapest, Konkoly Thege Miklós u. 29-33., Hungary; \\
	$^3$Department of Energy Engineering, Faculty of Mechanical Engineering,  Budapest University of Technology and Economics, H-1111 Budapest, Műegyetem rkp. 3., Hungary\\
	$^4$Montavid Thermodynamic Research Group, Budapest}
	\date{\today}

\begin{abstract}
The rotational velocity curve (RC) of galaxy NGC 3198 is modelled in various theoretical frameworks: Thermodynamic Gravity (TG) is compared to Dark Matter (DM) and Modified Newtonian Dynamics (MOND). The nonlinear gravitational field equation of TG is solved using the baryonic mass density as the source of the gravitational field. In this paper, first, a dissipation motivated numerical method is verified with the help of exact solutions. Then, the obtained optimal velocity curve is compared to \textcolor{black}{DC14 model} DM and \textcolor{black}{MOND EFE} \textcolor{black}{(Modified Newtonian Dynamics - External Field Effect aspect)} parametrisations for the RC of the galaxy. 
\end{abstract}

\maketitle

\section{Introduction}
	\setlength{\parindent}{2em}
	\setlength{\parskip}{1em}
	
The most successful model in cosmology so far is the so-called $\Lambda$CDM model, where $\Lambda$ stands for the cosmological constant in the Einstein-equation accounting for dark energy, and CDM stands for cold dark matter. The composition of DM is unknown. There is an ongoing theoretical, experimental and computational hunt to find proper candidates in the framework of existing or new particles of the Standard Model of Particle Physics \cite{Bertone_2018,x17,neutrinomass}.
		
The most well-known effect of dark matter is its influence on the rotation curves of galaxies. Properly modelling the observed galactic rotational velocity curves with DM requires particular dark matter distributions, \textcolor{black}{\cite{Famaey2012, Bertone_2018, KatzEtA17,SPARC_DM_over_MOND}}. However, in spite of the flexibility in modelling with the help of a potentially unknown distribution, there are some well-known problems and inconsistencies both in the case of individual galaxies and general, global properties.
	
Regularities in baryon-DM coupling can be summarised in the Radial Acceleration Relation (RAR). This relation generalises the more well-known dynamical relations, like Renzo's rule or the baryonic Tully-Fisher relation. The RAR encompasses these effects for various types of galaxies and establishes the strong connection between the baryonic contribution and the rotation curve \cite{Lelli_2017, Li2018}. These regularities indicate that DM is not a simple passive background but interacts with the baryonic mass component.
	
Related well-documented problems of the $\Lambda$CDM paradigm are the core-cusp problem, the missing satellite problem and the too-big-to-fail problem. The simulations that include dark matter predict that DM halo density profiles should rise steeply at small radii as $\rho(r) \propto r^{-\gamma}$ with $\gamma \simeq 0.8 - 1.4$ for small galaxies. This is in contrast to observations with many low-mass DM-dominated galaxies. A related 'central density problem' also exists as simulations predict more dark matter in the central regions than they should host. These two issues (low-density cores and high-density cusps) are summarised as the core-cusp problem \textcolor{black}{\cite{lambdacdm-challenges, Bol21a}}. That is one of the reasons that \textcolor{black}{hydrodynamics-motivated DM distributions are} widely used for rotational velocity curve modelling.
	
Because of the above-mentioned problematic aspects, there are several other proposals to explain the missing mass. The superfluid theories suggest a particular multi-component form of dark matter, \textcolor{black}{\cite{BerKho15a,covariant-emergent-grav, Zeilinger23, SPARC_DM_over_MOND, Mina22}}, where the explanation of rotational curves may come from a logarithmic field potential, \cite{Zlo22a, Sco22a}, or scale invariance, \cite{MaeGue20a1}.  
	
Among these suggestions, the most popular and most developed is MOND. Its success is based on the insight regarding the nonrelativistic, phenomenological observations, in particular, the galactic rotational curves. The reason became clear after the development of the first modified gravity version of MOND, called \textit{A QUAdratic Lagrangian theory} (AQUAL), \cite{aqual}. It has a vacuum solution with a force that is inversely proportional to the distance of the central mass: that can exactly counterbalance the centrifugal force and explain the typical flat velocity profiles. 

\textcolor{black}{While MOND has remarkable success in explaining many observational features of the local Universe \cite{refId0}, and TeVeS, the relativistic extension of AQUAL, is successful in explaining gravitational lensing and matter perturbation evolution, as Dodelson \cite{dodelson} points out, it still faces serious challenges, notably in the explanation of galaxy clusters, and the shape of the power spectrum of baryon-acoustic oscillations.}
 
A different approach was proposed by Verlinde, whose emergent gravity (also known as entropic gravity) uses the idea that spacetime and gravity emerge together from the entanglement structure of an underlying microscopic theory. Using insights from black hole thermodynamics, quantum information theory and string theory, emergent gravity is motivated and borrows elements of various nonrelativistic constitutive theories. For the DM phenomena, emergent gravity is motivated by generalised nonrelativistic elasticity \cite{Ver17a}. The AQUAL type field equation in the nonrelativistic limit can be obtained from a covariant version of the theory, \cite{Hos17a}.  
	
There is also a theory of gravity based on nonequilibrium thermodynamics, refered as thermodynamic gravity (TG) in the following, \cite{vanabe,AbeVan22a}. Then the  gravitational potential is treated as a thermodynamic state variable and the standard Newtonian gravity can be derived. The approach is based on the simple and only assumption that field energy and interaction energy is included in the total energy density. Then the inequality describing the entropy production rate in a local equilibrium state leads to the following dissipative field equation for gravity:
	\begin{equation}
		\frac{\partial \varphi}{\partial t} = \frac{l^2}{\tau} \left( \Delta \varphi - K \left( \nabla \varphi \right)^2 - 4 \pi G \rho\right),
		\label{dissipative-field-eq}
	\end{equation}
where $\varphi$ is the gravitational potential field, $l$ refers to the macroscopic spatial scale of variations of the whole system, $\tau$ is the relaxation time, $\Delta$ is the Laplace operator, $K$ is a characteristic coefficient of the dimensions $s^2m^{-2}$, $G$ is the Newtonian constant of gravity and $\rho$ is the mass density. The stationary solution yields a nonlinear field equation for a modified gravitational interaction, while the strength of deviation from the classical case is characterised by the $K$ constant, which arises from the linear coupling between the gravitational and mechanical interactions. This leads to a dynamic crossover between Newtonian and non-Newtonian regimes \textcolor{black}{in the vacuum region}. 
	
There, the relaxation to a stationary nonlinear field equation indicates the dissipative character of the theory. The nonlinear term is due to the presence of the gravitational field pressure, which is a consequence of the holographic properties of Newtonian gravity, and in general, the result of thermodynamic requirements, \textcolor{black}{\cite{Van23a, Logotropic_Chavanis22}}.
	
Therefore, due to the pressure of the gravitational field, there is a cross effect between the thermodynamic force of the scalar gravitational interaction and the scalar part of the mechanical interaction and the nonlinear term of eq. \ref{dissipative-field-eq} emerges naturally. Therefore, the parameter $K$ is not fundamental, it belongs to a particular transport property of the thermodynamic cross effect of gravitational field and mechanical momentum transport: the gravitational field transports material momentum and material momentum pulls the related gravity field along. {\color{black}$K$ is a material parameter that characterises the intensity of the aforementioned coupling in a galactic environment, therefore it is not a self-energy term as in the similar field equations of \cite{Giu97a,Sivaram2020,rebecca23}.} Also, it has a direct physical interpretation, as $v_K = K^{-1/2}$ is a characteristic velocity of a flat rotational curve because $F_K = \frac{1}{Kr}$ is the force field in a stationary vacuum solution of the field equation in TG.  
	
In the TG theory, the source term of the field equation is the density distribution of the baryonic mass, the sum of the visible matter and the atomic hydrogen densities. The gravitational field and the corresponding centrifugal force can be calculated by solving the stationary part of eq. \ref{dissipative-field-eq}. This paper serves as a proof of concept for this method, to be detailed for additional galaxies in following publications.
	
In the following sections, we give a finite difference scheme to solve the field equations and verify the method with the help of known exact solutions of the equation with constant, $\rho_0 \geq 0$ density distributions. Then, we demonstrate the applicability of the method, solving the equation with a realistic example, with the mass distribution of galaxy NGC 3198 as a source term. Here the sum of the visible part (referenced also as a stellar disk) and the atomic H (HI) component is used to solve the stationary part of the field equation \ref{dissipative-field-eq}. The numerical method is of a relaxational type, exploiting that the time-dependent field equation relaxes to the stationary solution.
	
The paper first presents the exact solutions of the equation. Then, a short description of the numerical method is given. Afterwards, we validate the \textcolor{black}{approach}, showing that the numerical scheme relaxes to the known exact solutions. Then, the velocity curve of the NGC 3198 galaxy is modelled, fitting the K parameter. Finally, a short analysis and interpretation follow.

\section{Thermodynamic gravity}

For the exact and numerical solutions, a spherically symmetric case will be treated, with the $r$ variable denoting the distance from the centre, as before. In this case, equation \ref{dissipative-field-eq} can be rewritten into the following form:
	\begin{equation}
		\partial_t \varphi = \frac{l^2}{\tau} \left( \partial_{rr} \varphi  + \frac{2}{r} \partial_r \varphi - 4 \pi G \rho - K (\partial_r \varphi )^2 \right).
		\label{sphere-tgrav}
	\end{equation}

\subsection{Exact solutions}
	
In this section the analytical solution for the stationary part of  eq. \ref{dissipative-field-eq} is investigated in the spherically symmetric case. Let us introduce the following shorthand notation:
	\begin{gather}
		\partial_r \varphi(r) =\varphi', \\
		\partial_{rr} \varphi(r) =\varphi''.
	\end{gather}
	\noindent
	If the field is time-independent, then $\varphi(\mathbf{r}, t)$ can be written for $\varphi(r)$ as
	\begin{equation}
		\varphi''  = -\frac{2}{r} \varphi' + 4 \pi 
		G \rho(r) + K \left(\varphi' \right)^2,
	\end{equation}
	\noindent
	whereby using gravitational acceleration, which is by definition
	\begin{equation}
		\mathbf{g} = - \nabla \varphi,
	\end{equation}
	\noindent
	and here becomes
	\begin{equation}
		g(r)= - \varphi',
	\end{equation}
	\noindent
	we may obtain the following form:
	\begin{equation}
		g(r)' = -\frac{2}{r} g(r) - 4 \pi G \rho(r) - K g(r)^2.
		\label{eq-for-g}
	\end{equation}
	
	This non-linear \textcolor{black}{Riccati} equation has well-known solutions when the density is zero or a constant value. 
 
	\subsubsection{Vacuum solution}
	
	For solutions with zero density, i.e. $\rho(r) = 0$, eq. \ref{eq-for-g} becomes
	\begin{equation}
		g(r)' = -\frac{2}{r} g(r)  - K g(r)^2,
		\label{eq-for-vacuum-g}
	\end{equation}
	\noindent
	which is a Bernoulli-type equation, and can be solved by introducing:
	\begin{equation}
		w(r) = g(r)^{-1},
		\label{eq-for-w}
	\end{equation}
	whereupon eq. \ref{eq-for-vacuum-g} becomes:
	\begin{gather}
		w(r)'-\frac{2}{r} w = K.
	\end{gather}
	\noindent
	This is a first-order linear ordinary differential equation, and has the following solution:
	\begin{equation}
		w(r) = - C r^2-K r.
	\end{equation}
	\noindent
	By substituting this into eq. \ref{eq-for-w}, one can obtain the following formulas:
	\begin{gather}
		g(r) = - \frac{1}{K r + C r^2}, \label{g-in-vacuum}\\
		\varphi (r) =  \frac{1}{K} \ln{\frac{r}{K + Cr}} + \varphi_{0}. 
	\end{gather}
	
This solution has a singularity at $r = 0$, so it may only be used after the effect of the mass distribution is negligible (e.g. far from the bulk of the galactic mass). The $C$ constant then may be fitted using the boundary condition at the transition to this regime. One can recover the Newtonian solution with a central point mass, if $K$ tends to zero:
	\begin{gather}
		-\frac{G M}{r^2} = g(r)_{K=0} =- \frac{1}{C r^2},\label{g-asymptotic-in-vacuum} \\
		C := \frac{1}{G M_{aa}}. \label{C-to-M}
	\end{gather}
	\noindent
This indicates that the parameter corresponds to an {\em apparent asymptotic equivalent point-mass} ($M_{aa}$) at the centre of our field. Therefore, introducing also the characteristic distance, $R = K/C$, one obtains the vacuum solution in the following intuitive form, \cite{AbeVan22a}:
    \begin{gather}
		g(r) = - \frac{G M_{aa}}{(R + r)r}. \label{reparam:g-in-vacuum}
    \end{gather}
 
\subsubsection{Solution with constant density}
	
	Eq. \ref{eq-for-g} can be solved analytically if the density is a nonzero $\rho_c$ constant, too. Then a general solution can be written as 
	\begin{equation}
		g(r) = \frac{1}{Kr} + \frac{\sqrt{- 4 \pi G \rho_c K}}{K} \tan \left( \sqrt{-4 \pi G \rho_c K} \cdot r + C \right).
		\label{general-const-rho-solution-for-g}
	\end{equation}
	
	For further consideration, let us examine the cases where $K > 0$ and $K < 0$. In the $K > 0$ case, we can rewrite eq. \ref{general-const-rho-solution-for-g} as 
	\begin{gather}
		g_{+}(r) = \frac{1}{Kr} + \sqrt{\frac{ 4 \pi G \rho_c}{ K}} \cdot i \tan \left(i \sqrt{4 \pi G \rho_c K} \cdot r + C \right), \\
		i\Tilde{C} = C, \\
		g_{+}(r) = \frac{1}{Kr} - \sqrt{\frac{ 4 \pi G \rho_c}{ K}} \cdot  \tanh \left( \sqrt{4 \pi G \rho_c K} \cdot r + \Tilde{C} \right),
	\end{gather}
	\noindent
	and in the $K < 0 $ case, as
	\begin{gather}
		g_{-}(r) = -\frac{1}{|K|r} - \sqrt{\frac{ 4 \pi G \rho_c}{ |K|}} \cdot  \tan \left( \sqrt{4 \pi G \rho_c |K|} \cdot r + C \right).
	\end{gather}
	
	Here $i$ is the imaginary unit, $C$ and $\tilde C$ are constants. Since for this constant density mass model we want to eliminate the singularity at $r= 0$, we may do so by appropriately choosing $C$ or $\Tilde{C}$. By choosing $\Tilde{C} = i \frac{\pi}{2}$, the $\tanh(x)$ function becomes $\coth (x)$, and will have the following series representation in the equation:
    \begin{gather}
		g_{+}(r) = \frac{1}{Kr} - \sqrt{\frac{ 4 \pi G \rho_c}{ K}} \cdot  \coth \left( \sqrt{4 \pi G \rho_c K} \cdot r  \right) 
		\label{g+_no-singularity}\\
		 \approx \frac{1}{Kr} - \sqrt{\frac{ 4 \pi G \rho_c}{ K}} \cdot  \left( \frac{1}{\sqrt{4 \pi G \rho_c K} \cdot r}  + \frac{\sqrt{4 \pi G \rho_c K} \cdot r}{3} - \frac{\left(\sqrt{4 \pi G \rho_c K}\right)^3 \cdot r^3}{45} + \mathcal{O}\left(r^5 \right) \right) \nonumber\\
		\approx -\frac13 \cdot 4 \pi G \rho_c  \cdot r + \frac{K}{45} \left(4 \pi G \rho_c \right)^2 \cdot r^3 + \mathcal{O}\left(r^5 \right), \nonumber
	\end{gather}
	\noindent
	and thus eliminates the singularity. We can note that the first term corresponds to the usual result in the case of Newtonian gravity, while the higher-order terms, which also contain the $K$ parameter, serve as corrections. For $g_{-}(r)$, we can choose $C = \frac{\pi}{2}$ and obtain similarly:
	\begin{gather}
            g_{-}(r) = -\frac{1}{|K|r} + \sqrt{\frac{ 4 \pi G \rho_c}{ |K|}} \cdot  \cot \left( \sqrt{4 \pi G \rho_c |K|} \cdot r  \right)
		\label{g-_no-singularity}\\
			\approx -\frac{1}{|K|r} + \sqrt{\frac{ 4 \pi G \rho_c}{ |K|}} \cdot   \left(\frac{1}{ \sqrt{4 \pi G \rho_c |K|} \cdot r} - \frac{\sqrt{4 \pi G \rho_c |K|} \cdot r}{3} - \frac{ \left(\sqrt{4 \pi G \rho_c |K|} \right)^3 \cdot r^3}{45} + \mathcal{O}(r^5)  \right) \nonumber\\
		\approx -\frac13 \cdot 4 \pi G \rho_c  \cdot r - \frac{|K|}{45} \left(4 \pi G \rho_c \right)^2 \cdot r^3 + \mathcal{O}\left(r^5 \right).
  \nonumber\end{gather}
\noindent
which is the same in the first two orders as in the previous case, considering the sign of $K$. The difference only appears in terms containing higher orders of $K$, as the two series diverge.

	\subsection{Numerical solution}
		
	This chapter details the numerical method for the solution of eq. \ref{sphere-tgrav}. Here, generally, $\rho(r, t)$ can be a function of time, as it is coupled to the fundamental hydrodynamic balances. However, changes in the density due to the characteristic speed of matter will be considered negligible relative to the relaxation of the potential. This is justified by experimental boundaries on the possible delay time of the gravitational field, summarised in the work of Diósi \cite{DIOSI20131782}, and the characteristic timescale must be below  millisecond. This insight also serves as a justification to numerically treat the gravitational potential as relaxing to the stationary solution in eq. \ref{sphere-tgrav}.

	\subsubsection{Dimensionless formalism}
	
		In order to solve the equation, a dimensionless form is created, by introducing the following notation:
	\begin{gather}
		\rho = \rho_d \tilde{\rho}, \\
		\frac{l^2}{\tau} = a, \\
		r = r_d \tilde{r}, \\
		\varphi = \varphi_d \tilde{\varphi}, \\
		K = \frac{1}{\varphi_d} \tilde{K}, \\
		t = t_d \tilde{t} \rightarrow t_d = \frac{r_d^2 \tau}{l^2} 
  = \frac{r_d^2}{a},
	\end{gather}
	\noindent
	where the quantities with the index $d$ denote a value with appropriate dimensions and the values with a tilde are dimensionless. Substituting these into eq. \ref{sphere-tgrav}, we obtain the following:
	\begin{gather}
		\frac{\partial \left( \varphi_d \tilde{\varphi} \right)}{\partial \left( t_d \tilde{t}\right)}  = a \left( \frac{\partial^2 \left(\varphi_d \tilde{\varphi} \right)}{\partial \left(r_d \tilde{r} \right)^2} + \frac{2}{r_d \tilde{r} } \frac{ \partial \left( \varphi_d \tilde{\varphi} \right) }{\partial \left(r_d \tilde{r} \right)} - 4 \pi G \rho_d \tilde{\rho} - \frac{1}{\varphi_d} \tilde{K} \left( \frac{ \partial \left( \varphi_d \tilde{\varphi} \right) }{\partial \left(r_d \tilde{r} \right)} \right)^2  \right), \\
		\frac{\partial  \tilde{\varphi}}{\partial  \tilde{t}}  =  \frac{\partial^2  \tilde{\varphi}}{\partial  \tilde{r}^2} + \frac{2}{\tilde{r} } \frac{ \partial  \tilde{\varphi}  }{\partial  \tilde{r}} - \frac{4 \pi G \rho_d r_d^2}{\varphi_d} \tilde{\rho} -  \tilde{K} \left( \frac{ \partial \tilde{\varphi}  }{\partial \tilde{r}} \right)^2 .
	\end{gather}
	\noindent    
	Let us also choose
	\begin{equation}
		\frac{4 \pi G \rho_d r_d^2}{\varphi_d}
		\overset{!}{=} 1.
	\end{equation}
	
	\noindent
	By leaving the tilde notation, we obtain the equation analogue to eq. \ref{sphere-tgrav} in the following form:
	\begin{equation}
		\partial_t \varphi   =  \partial_{rr} \varphi + \frac{2}{r} \partial_r \varphi -\rho -K \left( \partial_r \varphi \right)^2.
		\label{clean-tgrav}
	\end{equation}
	
	For the numerical method, $r_d$ is chosen to be the furthest point of the velocity curve data and $\rho_d$ as the mean of the derived density distribution, as can be seen later in eq. \ref{rd}.
	
	\subsubsection{Discretisation}
	
	In order to solve eq. \ref{clean-tgrav} numerically, a discretisation method of staggered grids is employed \cite{nummodszerek,RieEta18a,PozsEta20a}. This means that the derivative of the field is introduced as $q =  \partial_r \varphi$, which allows us to rewrite eq. \ref{clean-tgrav} as a system of two first-order differential equations:
	\begin{gather}
		\partial_t \varphi   =  \partial_{r} q +\frac{2}{r}q-\rho-K q^2, \\
		q =  \partial_r \varphi.
	\end{gather}

\begin{figure}[h]
		\centering
		\includegraphics[width=0.8\textwidth]{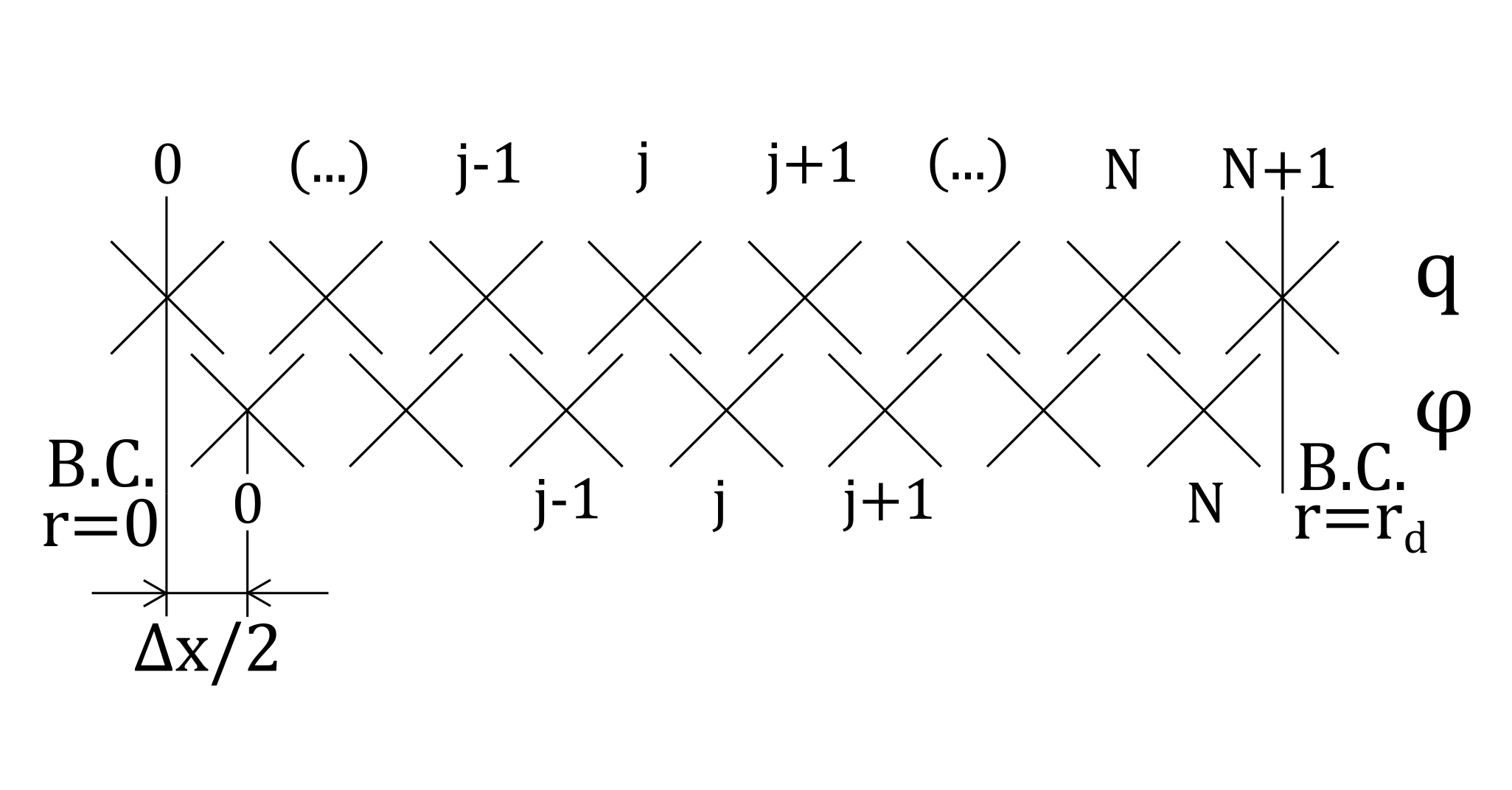}
		\caption{The schematic representation of the discretisation method of staggered grids, which was used in the numerical solution \cite{nummodszerek}. The (Neumman-type) boundary conditions (B.C.) are set for the derivative field.}
		\label{fig:num-schema}
	\end{figure}

	The aim of this method is the numerical prescription of Neumann-type boundary conditions and the better numerical handling of the nonlinear term with the coefficient $K$. During the numerical solution, in a given timestep, the $\varphi$ field is evaluated first, using the values of $\varphi$ and $q$ from the previous timestep, and then the derivate field $q$ is evaluated, using the obtained new field value. In the staggered grids method, the two variables are shifted by $\Delta x /2 $ for better handling. The schematic representation of this can be seen in Fig. \ref{fig:num-schema}. Henceforth, $r$ will be shown with dimensional values (\textit{e.g.} $r = r_d$ for $\tilde{r} = 1$) for easier comprehension, however, all other physical quantities remain dimensionless.
	
	For the terms containing $q$ explicitly, the average of the $q$ field at two spatially neighbouring points can be used. By denoting the spatial indices with $j$ and the temporal indices with $t$, the two equations can be simply discretised in the following way:
	
	\begin{gather}
		\varphi_{j, t+1} = \varphi_{j, t} + \Delta t \left(\frac{q_{j+1, t}-q_{j, t}}{\Delta x} + \frac{2}{\left( \frac12 + j \right) \cdot \Delta x} \frac{q_{j,t}+q_{j+1,t}}{2} - \rho_j - K \left( \frac{q_{j, t} +q_{j+1, t}}{2} \right)^2\right), \label{fi-discreet} \\
		q_{j, t+1} = \frac{\varphi_{j, t+1} - \varphi_{j-1, t+1}}{\Delta x}.
		\label{q-discreet}
	\end{gather}
	
	In order to ensure the stability of the numerical method, a sufficiently small $\frac{\Delta t}{\Delta x^2}$ ratio was chosen.  Unless otherwise noted, for the numerical calculations, $N = 200$ was used.
	The initial conditions for the arrays $\varphi$ and $q$ were chosen to be
	\begin{gather}
		\varphi(t=0) = 0, \\
		q(t= 0) = 0,
	\end{gather}
	\noindent
	except in the boundaries, while the boundary conditions were\textcolor{black}{
	\begin{gather}
		 q(r = r_{min}) = \frac{v_{observed}(r_{min})^2}{\varphi_d},   
   \end{gather}}
   \begin{gather}
   		q(r=r_d) = \frac{v_{observed}(r_d)^2}{\varphi_d}, 
	\end{gather}
	\noindent
during the galaxy velocity curve calculation\textcolor{black}{, with $r_{min}$ denoting the radial position of the innermost observed velocity data point.} For the reliability testing, the boundaries were set to the corresponding values of the theoretical curves.
	
	\subsubsection{Convergence and reliability}
	
	In order for the numerical solution to yield a good enough approximation of the stationary solution, the number of the timesteps needs to be sufficiently high. To ensure this, the number of temporal iterations $T$ is set to fulfil the following:
	\begin{equation}
		T \cdot \Delta t = 1 \rightarrow T = \frac{1}{\Delta t}.
	\end{equation}
	
\begin{figure}[h]
		\centering
		\includegraphics[width=0.8\textwidth]{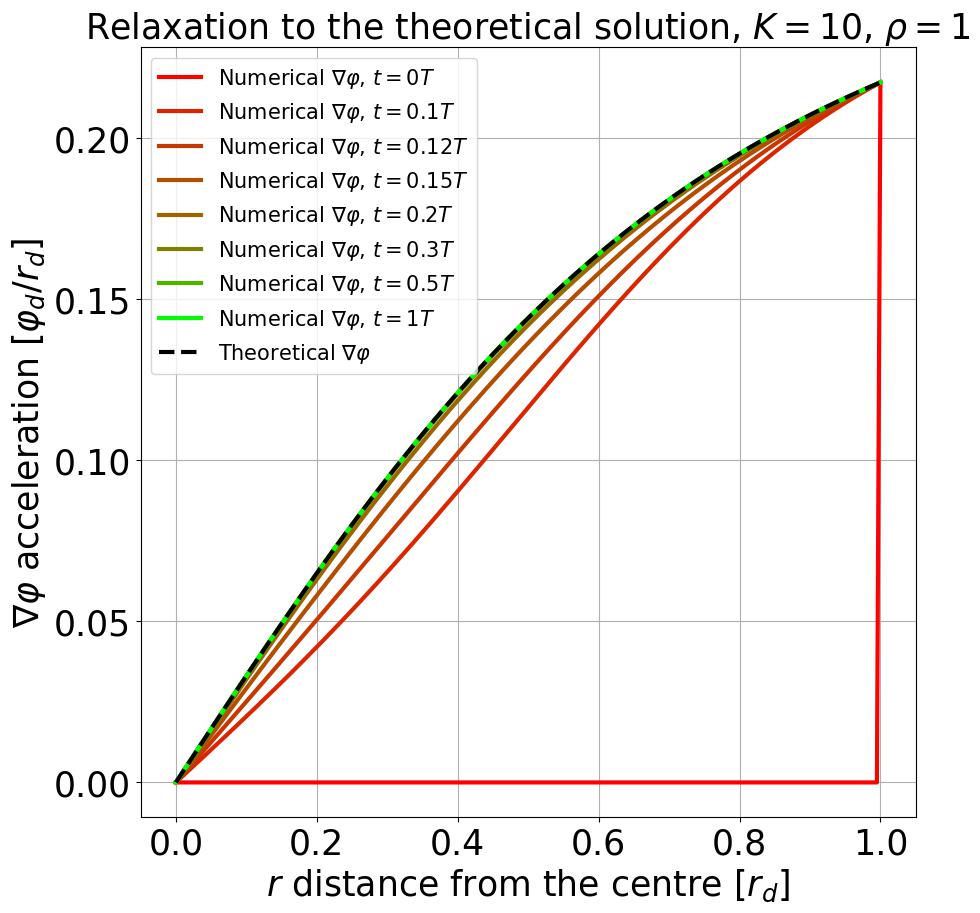}
		\caption{The numerical method shown to be relaxing to the stationary solution in the $\rho(r) = \rho_0$ solution, with $N= 200$ and dimensionless values of $K = 10$ and $\rho = 1$.}
		\label{fig:relaxation}
	\end{figure}

	The relaxation process is shown for the relaxation of the potential and the derivative in Fig. \ref{fig:relaxation}. As can be seen, the method converges fast, and this choice provides enough timesteps for a good convergence to the theoretical solution.
	
\begin{figure}[h!]
		\centering
		\includegraphics[width=0.4\textwidth]{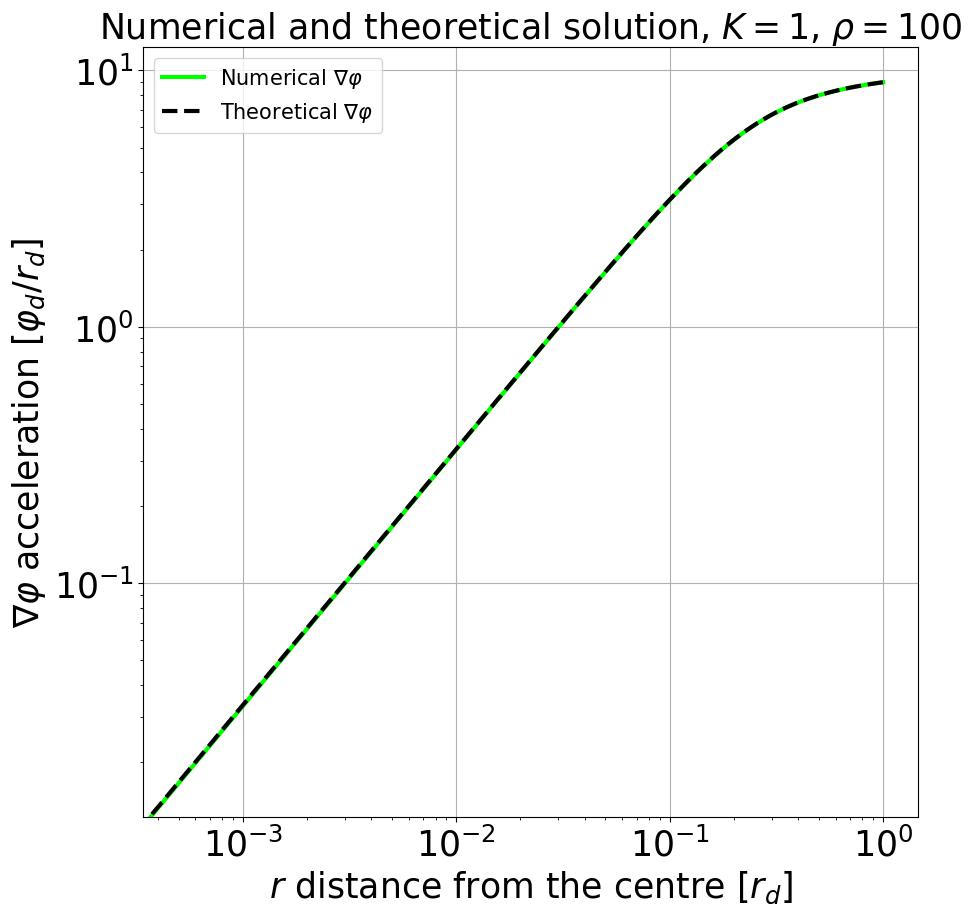}
		\includegraphics[width=0.4\textwidth]{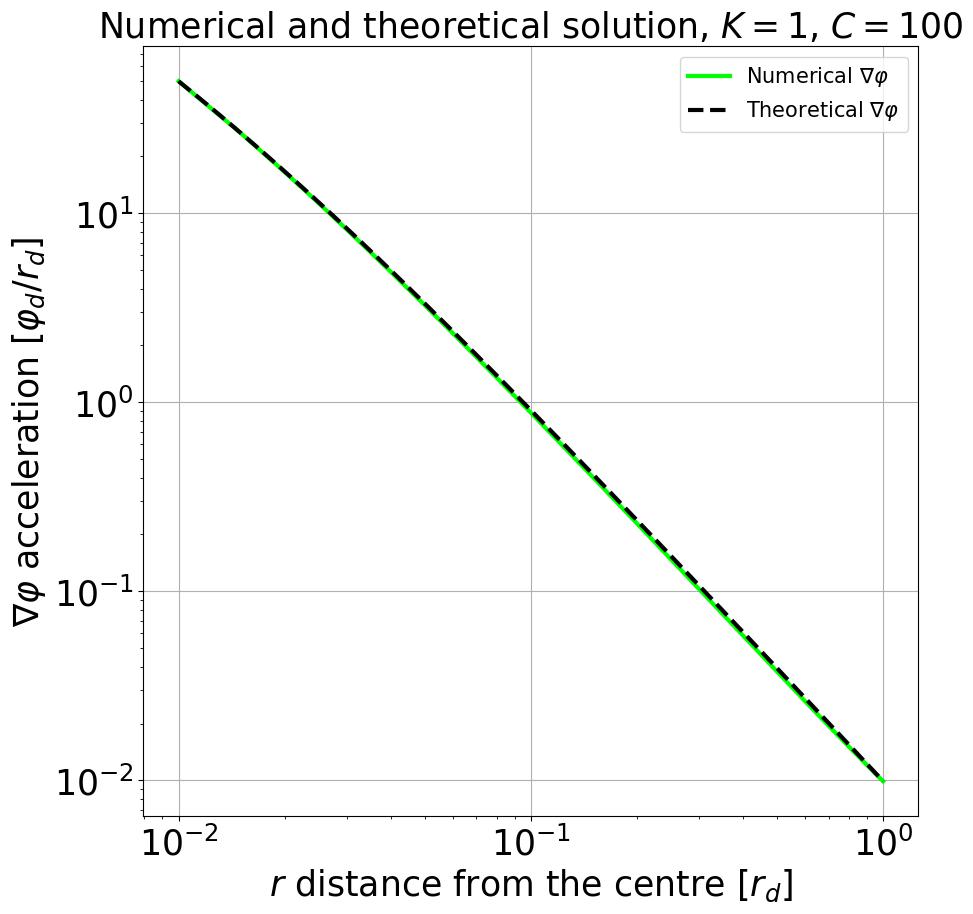}
		
		\includegraphics[width=0.4\textwidth]{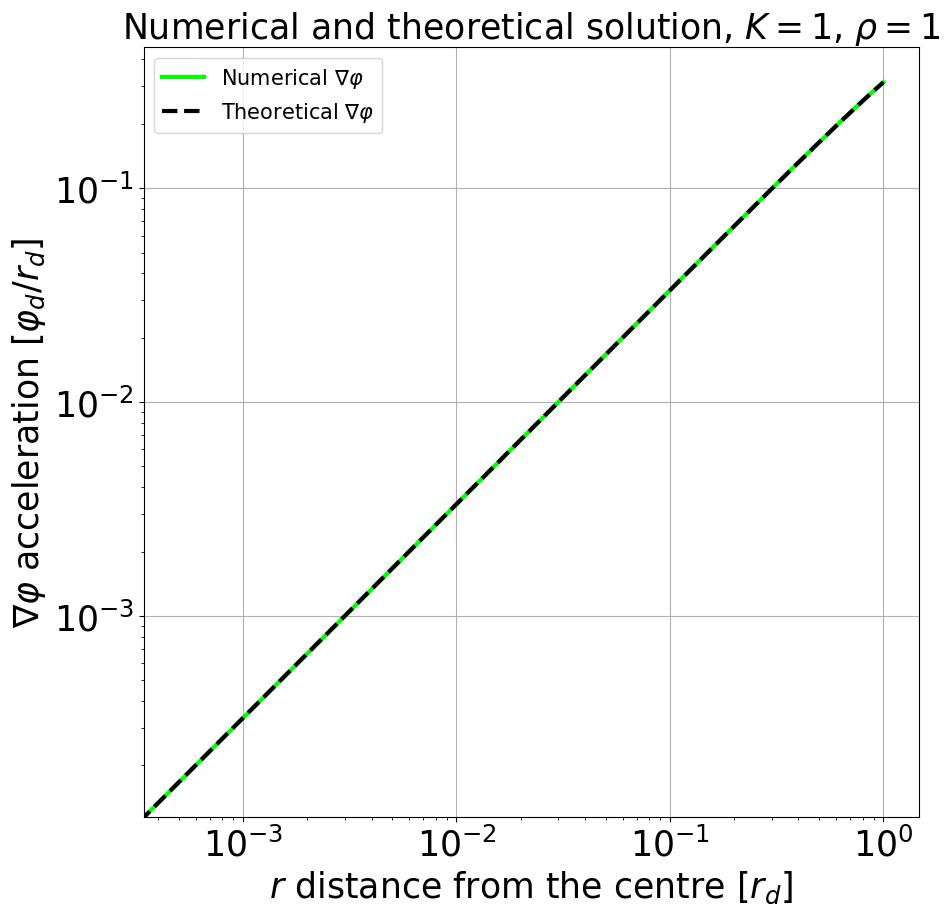}
		\includegraphics[width=0.4\textwidth]{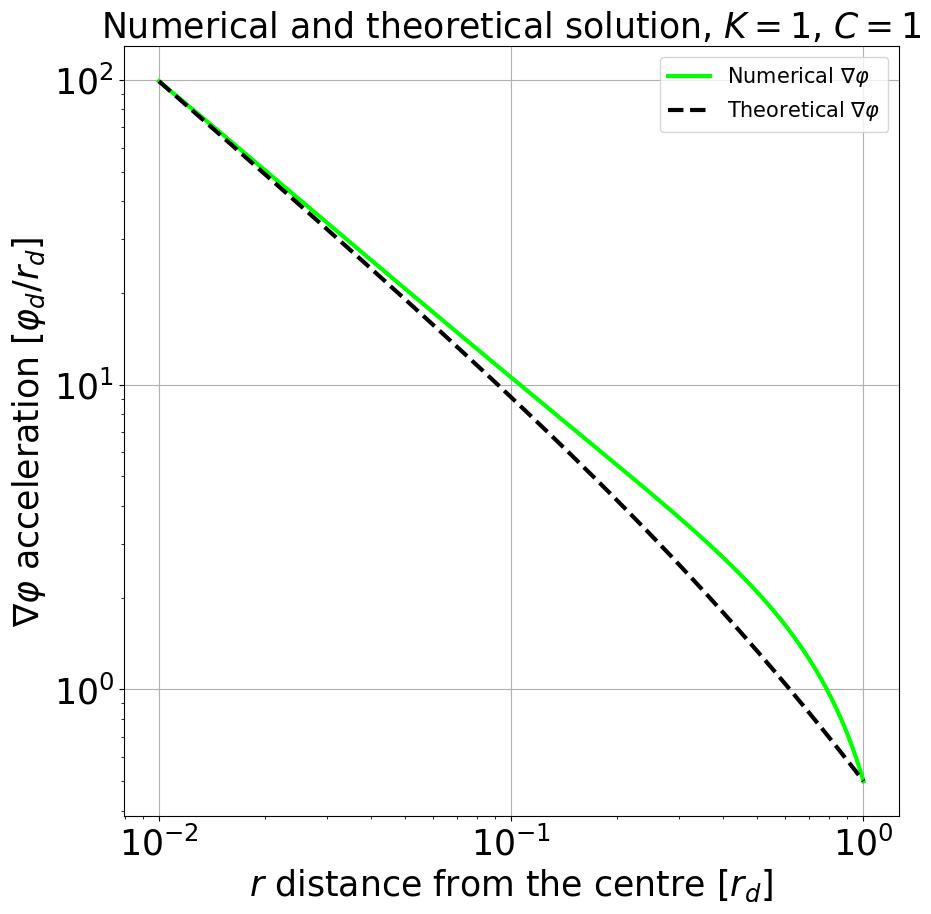}
		
		\includegraphics[width=0.4\textwidth]{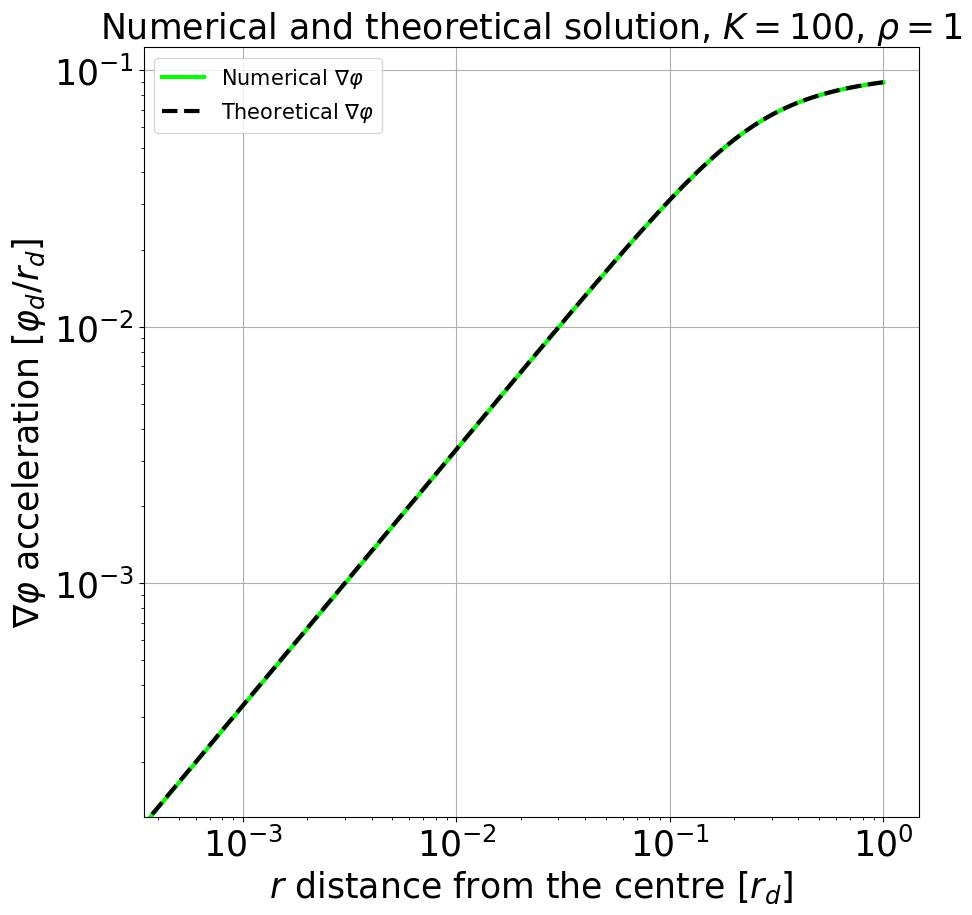}
		\includegraphics[width=0.4\textwidth]{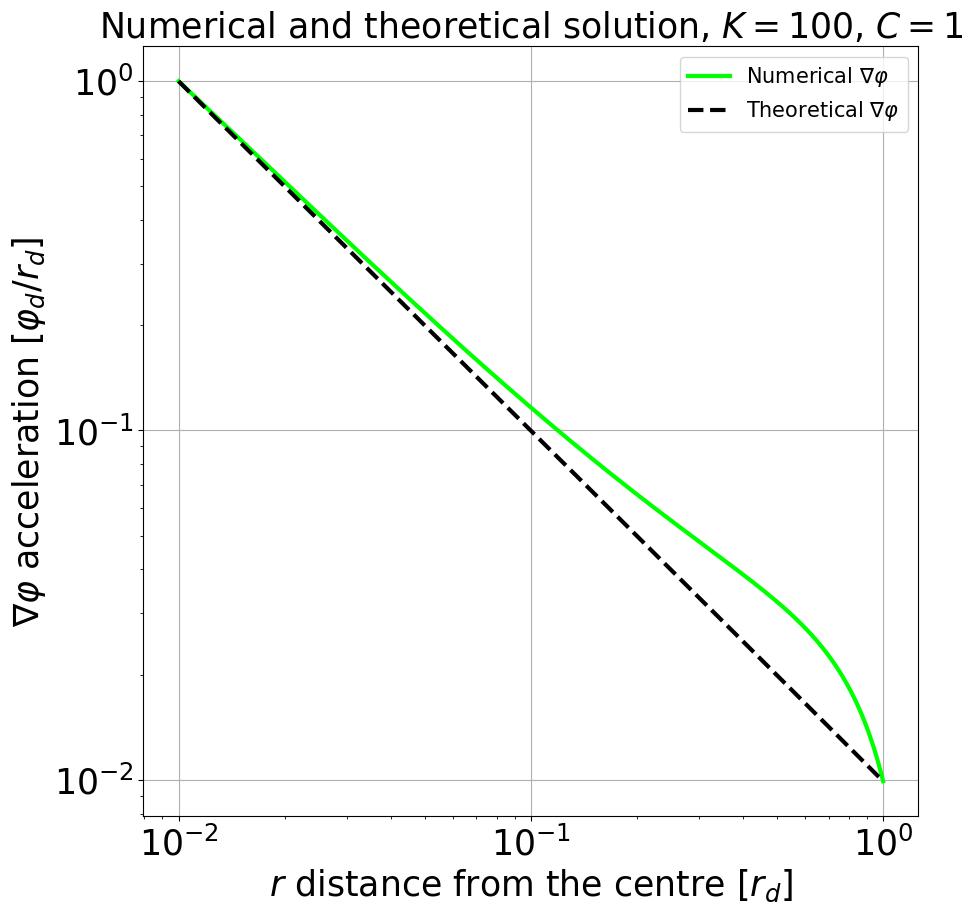}
		
		\caption{The result of the numerical method compared to the theoretical solution in the regimes of $K \ll \rho_0$, $K = \rho_0$ and $K \gg \rho_0$ for constant density and for $K \ll C$, $K = C$ and $K \gg C$ for vacuum.}
		\label{fig:num-vs-theory}
	\end{figure}

	To test the reliability of the numerical method, its results are compared to the theoretical results from eqs. \ref{g-in-vacuum}, \ref{g-_no-singularity} and \ref{g+_no-singularity}. Note that due to the nonlinear term, the numerical method and the theoretical solutions can only be compared with an opposite sign of $K$. The results are shown in Fig. \ref{fig:num-vs-theory}.
%	\newpage
	
\begin{figure}[h]
		\centering
		\includegraphics[width=0.8\textwidth]{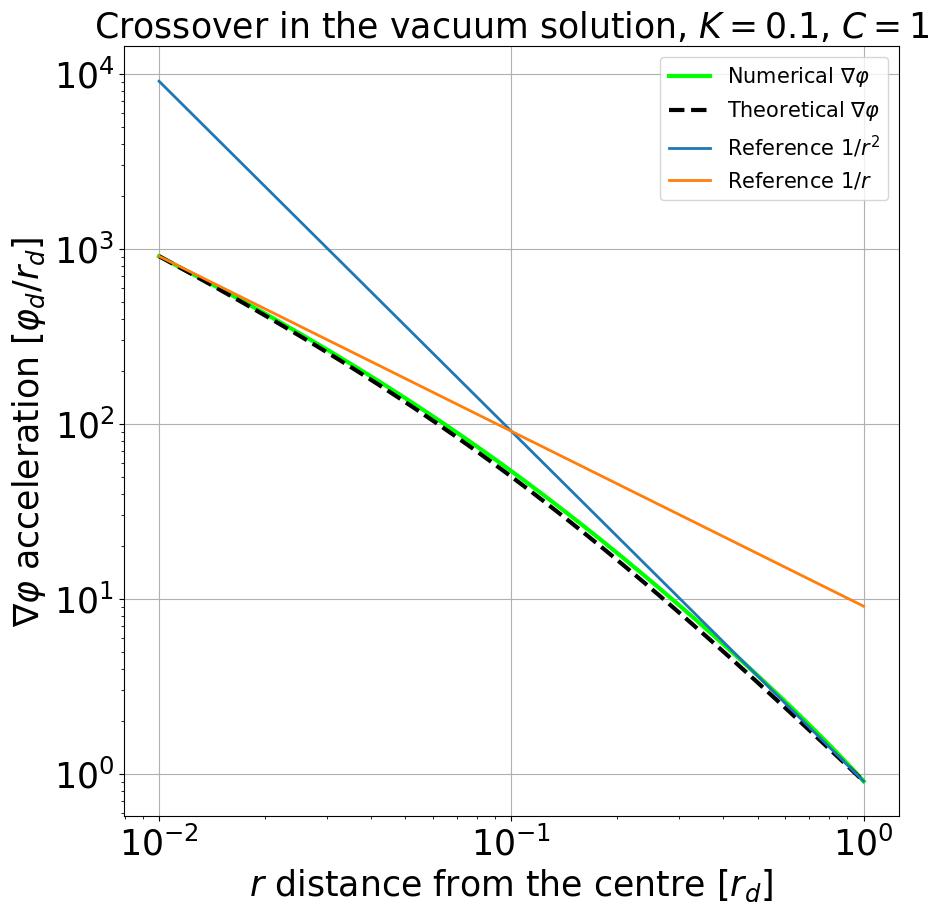}
		\caption{The crossover in the vacuum solution. Near the inner boundary, the $\frac{1}{Kr}$ term dominates, and at the outer boundary, the $\frac{1}{Cr^2}$ dominates. The resulting field looks Newtonian from far away, but with a greater mass, compared to the classical case.}
		\label{fig:double-crossover}
	\end{figure}

	We can see that the numerical solution reproduces the exact theoretical solution for constant density, but only reproduces the theoretical solution closer to the boundary for vacuum in the $K \ll C$ regime, that is, when the effect of the apparent mass is much greater than the effect of the nonlinear term.  
 By choosing the appropriate scale, we can see the crossover as in \cite{AbeVan22a} on Fig. \ref{fig:double-crossover}.
	
	Furthermore, the joint solution of a nonzero density core surrounded with vacuum is tested, where $\rho(r) = \rho_c$ in the interval $r \in [0,\frac12 r_d )$, and $\rho(r) = 0$ in the interval $r \in [\frac12 r_d, r_d]$. To determine the $C$ parameter in the vacuum solution, the continuity of the derivative was prescribed:
	\begin{gather}
		g_{\pm}\left(\frac12 r_d \right) = g_{vacuum} \left(\frac12 r_d\right), \\
		C = -\frac{4}{g_{\pm}\left(\frac12 r_d \right) r_d^2} - \frac{2K}{r_d}.
	\label{Cjoint}\end{gather}
	
\begin{figure}[h!]
		\centering
		\includegraphics[width=0.45\textwidth]{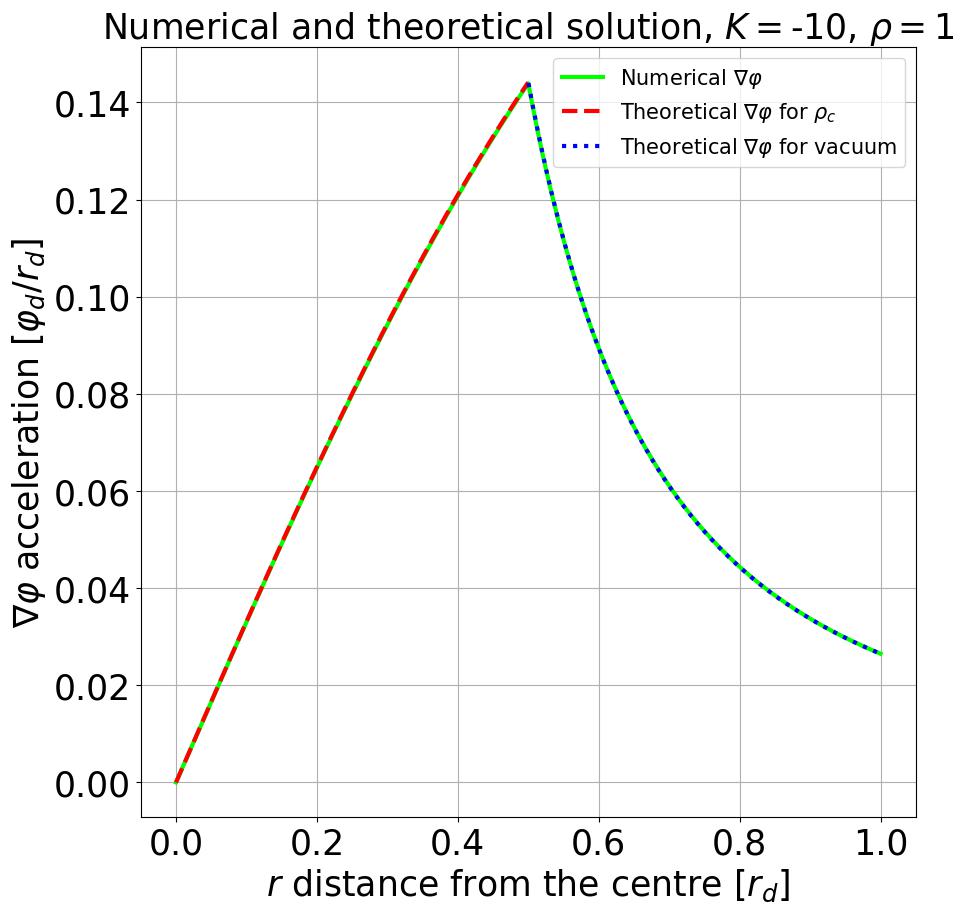}
		\includegraphics[width=0.45\textwidth]{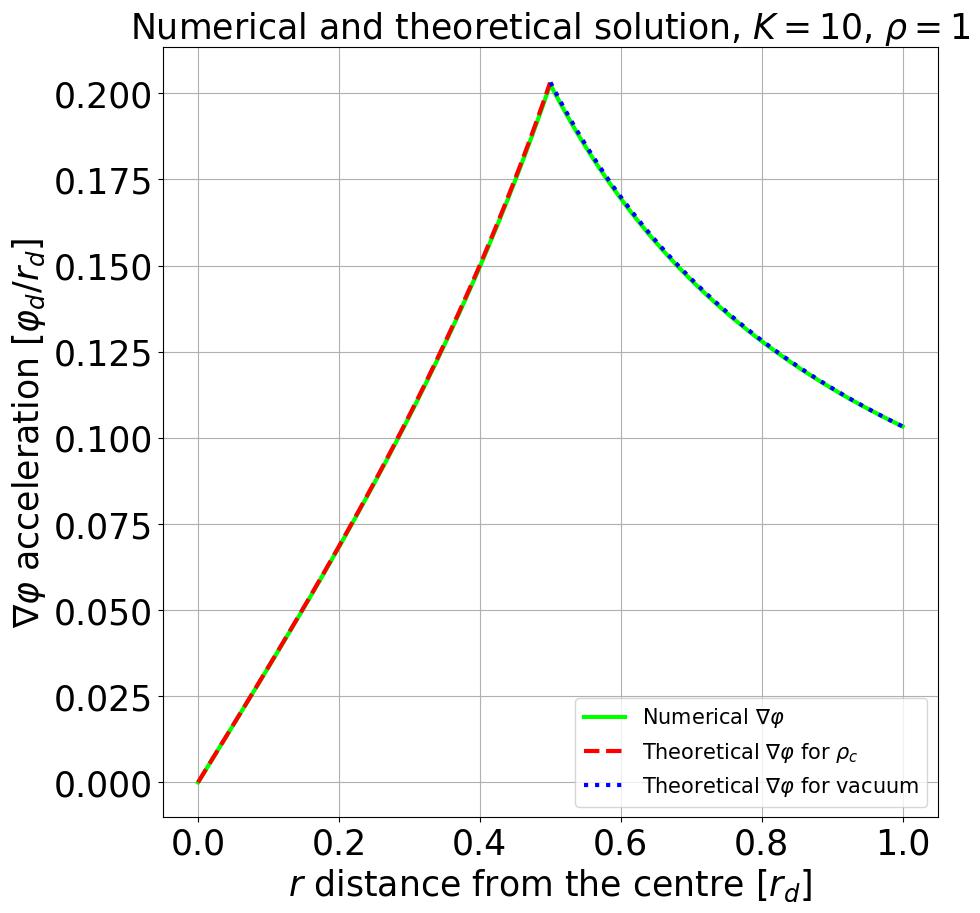}
		
		\caption{The result of the numerical method compared to the joint theoretical solutions in the case of dimensionless values $\rho_c = 1$, $K = - 10$ and $K = 10$.}
		\label{fig:num-vs-theory-joined}
	\end{figure}

	The boundary conditions are the corresponding constant density and vacuum derivatives. The numerical method is shown to reproduce the theoretical joint case in of dimensionless values $\rho_c = 1$,  $K = -10$ and $K = 10$ in Fig. \ref{fig:num-vs-theory-joined}. 

\newpage

Therefore we can trust that the numerical solution for the thermodynamically modified gravitational field will be reliable for realistic mass distributions as well, as long as they have similar characteristic parameter relations. Larger $K$ values relative to $\rho_c$ result in negative $C$ as per eq. \ref{Cjoint} and are unlikely to be physically meaningful as per eq. \ref{C-to-M}.
 
%	\newpage
	\section{Analyses of rotational velocity curves}
	
	\subsection{Methodology
	}
	To calculate velocity curves, the derivative field from the numerical solution is used to calculate the inferred rotational velocity, calculating the dimensional values:
	\begin{equation}
		v_j = \left(\frac{\varphi_d}{r_d}\;  r\; q_{j,T} \right)^{\frac12}.
	\end{equation}
 
	The velocities corresponding to the surface densities for atomic hydrogen (HI) and the stellar disk (SD) were converted into an equivalent pseudo-spherical dataset to calculate density data. The original velocities are defined as the velocities each component would induce in the plane of the galaxy, and for the comparison a spherical case is considered.
 \textcolor{black}{The Spitzer Photometry and Accurate Rotation Curves (SPARC) database (Lelli et al. \cite{SPARC_master})} uses the convention that the negative sign of the \textcolor{black}{hydrogen gas} data means that it has a negative contribution when calculating the total $v^2$ sum, that is, it has an outward gravitational effect due to the lack of spherical symmetry and the specifics of its distribution. The observed velocity curve data from SPARC is based on \textcolor{black}{Spitzer photometry at 3.6 $\mu $m (tracing the stellar mass distribution) and high-quality HI+H$\alpha$ rotation curves.}
 To calculate the $\rho_j$ for the numerical method from the interpolated function of the rotation velocities $v_{rot}(r)$, the following procedure was used:
	\begin{equation}
		\rho_j   = \mathcolor{black}{ \Upsilon_1} \cdot \left[\frac{  \partial (v_{rot, SD}(r)^2 r )}{\partial r} \right]_j \cdot \frac{1}{4 \pi G r_j^2} + \left[ \frac{ \partial (v_{rot, HI}(r)^2 r )}{\partial r} \right]_j \cdot \frac{1}{4 \pi G r_j^2},
		\label{pseudo-sphere-rho}	\end{equation}	\noindent
	\textcolor{black}{with $\Upsilon_1 = M_*/L$ denoting the fitted mass-luminosity ratio and} $\rho_d$ was selected to be the mean of the array $\rho_j$:
	\begin{gather}
		\rho_d = \mathrm{mean}(\rho_j), \label{rd} \\
		\tilde{\rho}_j = \frac{\rho_j}{\mathrm{mean}(\rho_j)}.
	\end{gather}

	For comparison, \textcolor{black}{ we show a dark matter and a MOND modelling parametrisation. 
 The dark matter model with DC14 density profile and with flat prior is used, from the work of Li et al. \cite{LiEta19a,  Li_2020}. This model has the density profile
 \begin{equation}
     \rho_{\mathrm{DC14}} (r)= \frac{\rho_s}{\left(\frac{r}{r_s} \right)^\gamma \left[1+ \left( \frac{r}{r_s}\right)^\alpha \right]^{(\beta - \gamma)/\alpha}},
 \end{equation}
 where $\rho_s$ is a characteristic volume density, $r_s$ is the scale radius; $\beta$ and $\gamma$ are the inner and outer slopes respectively, and $\alpha$ describes the transition between the inner and outer regions. The last three paramaters' value depends on the stellar-to halo mass ratio (adopted from \cite{LiEta19a}), and ultimately, the corresponding dark matter rotation curve is given by
 \begin{equation}
     v_{\mathrm{DC14}}(r) = v_{200} \sqrt{\frac{C_{200}}{x} \frac{ B(a,b+1, \epsilon) +B(a+1,b, \epsilon)}{B(a,b+1, \epsilon_c) + B(a+1,b, \epsilon_c)}},
 \end{equation}
 with $a = (3-\gamma)/\alpha$, $b =(\beta-3)/\alpha$, $\epsilon= \frac{(r/r_s)^\alpha}{1+ (r/r_s)^\alpha}$, $\epsilon_c = \frac{C_{200}^\alpha}{1+C_{200}^\alpha}$, $B(a, b, x)$ being the incomplete beta-function and $x = \frac{r}{r_s}$. The above formula is obtained after calculating the total mass of the DC14 mass distribution inside a given radius $r$. The $C_{200}$ and $v_{200}$ parameters were used in their fitting method, and the relate to the fitted physical parameters as
 \begin{equation}
     C_{200} = r_{200}/r_s, ~v_{200} = 10 C_{200} r_s H_0, 
 \end{equation}
 with $r_{200}$ being the virial radius inside of which the average density is 200 times the critical density of the Universe ($\rho_{\mathrm{crit}} = \frac{3 H_0^2}{8 \pi G}$) and $H_0 = 73 \mathrm{kms}^{-1} \mathrm{Mpc}^{-1}$ is the Hubble constant.
 For the galactic calculation, the fitted parameters for the flat prior DC14 dark matter model are taken from the SPARC website\footnote{\hyperlink{asd}{http://astroweb.cwru.edu/SPARC}} from \cite{LiEta19a, Li_2020}. To obtain the velocity curve from the DC14}  DM model, the contributions of the SD and HI are added to this:
\begin{equation}
    v_{\mathrm{DM}} (r) = \sqrt{v_{\mathrm{DC14}}^2(r) + (M_*/L) v_{rot, SD}(r)^2 + v_{rot, HI}(r) \cdot \mathrm{abs}\left({v_{rot,HI}(r)}\right)}.
\end{equation}
\textcolor{black}{Regarding MOND}, the method and data of \textcolor{black}{Chae et al. are used, \cite{Chae_2021}}.
  The exact form of the interpolating function of MOND is not fixed by the theory itself and some forms work better in explaining certain phenomena. {\textcolor{black}Zhao \& Famaey \cite{Zhao_2006} found that a simplified form of the interpolating function not only provides good fits to the observed rotational curves but also the derived mass/luminosity ratios are more compatible with those obtained from stellar populations synthesis models. It is also important to remark, that considering the observation uncertainties, like with the mass/luminosity ratio, not only the quality of the modelling is improved, but also there is no need to deviate from the standard MOND acceleration, $a_0$, contrary to the modelling approach of Wang and Chen \cite{Wang_2021}.  \textcolor{black}{In \cite{Chae_2021}} \textcolor{black}{the expected acceleration of a test particle ($v^2/r$ for circular motion) is given by \begin{equation}
     g = \nu_{e_N} \left( y \equiv g_N/a_0 \right) g_N,
 \end{equation}
 with $g_N$ as the Newtonian internal gravitational field (from the galactic matter), $a_0 = 1.2 \cdot 10^{-10} \mathrm{m s}^{-2}$ MOND acceleration scale and $\nu_{e_N}$ being
\begin{equation}
    \nu_{e_N}(y) = \frac12 +\sqrt{\frac14 D^2_{e_N}(y) + \frac{ D_{e_N}(y)}{y}} - \frac{e_N}{\sqrt{|e_N|}}\frac{C_{e_N}}{y},
\end{equation}
 where $C_{e_N} \equiv \sqrt{1+ |e_N|/4}$, $D_{e_N}(y) \equiv 1+ |e_N|/y$ and the fitted parameter is $\tilde{e} \equiv e_N/ \sqrt{|e_N|}$.
 Here, $g_N$ is given as
 \begin{equation}
     g_N = \frac{v_{\mathrm{bar}}^2}{r},
 \end{equation}
 with\textcolor{black}{
\begin{equation}
    v_{\mathrm{bar}}(r) = \sqrt{(M_{*, \mathrm{disk}}/L) v_{rot, SD}(r)^2 + (M_{ \mathrm{gas}}/M_{\mathrm{HI}}) v_{rot, HI}(r) \cdot \mathrm{abs}\left({v_{rot,HI}(r)}\right)}
\end{equation}
}
   describing the contribution of the stellar disk and the neutral hydrogen. 
Both the DM and MOND EFE methods fit the distance to the galaxy and its inclination, which can be summarised as
\begin{gather}
    v_{\mathrm{bar, fit}}^2 = \frac{D_{\mathrm{fit}}}{D_{\mathrm{obs}}}v_{\mathrm{bar}}^2, \\
    v_{\mathrm{obs, fit}} = v_{\mathrm{obs}} \frac{\sin{i_{\mathrm{obs}}}}{\sin{i_{\mathrm{fit}}}}, \\
    r_{\mathrm{fit}} = \frac{D_{\mathrm{fit}}}{D_{\mathrm{obs}}} r_{\mathrm{obs}}
\end{gather}
with $v_{\mathrm{obs}}$ denoting the observed rotational velocity of the galaxy and $v_{\mathrm{obs, fit}}$ its fit-adjusted value. In our method, this gives diminishing return, thus for the better quality of the algorithmic fit, we omit these parameters, but note that their uncertainties are important to consider when evaluating galactic RC fits.}

	\subsection{Galactic rotational velocity curve of NGC 3198}
	
	In this section, the above-mentioned method is applied to reproduce the velocity curve of NGC 3198. \textcolor{black}{The parameters $\Upsilon_1$ and $K$ were fitted using \texttt{scipy} and their presented error is the resulting estimate from the fit. The reduced chi-squared value is calculated as the following:
 \begin{equation}
\chi^2_\nu = \sum_r \frac{\left( v_{obs}-v_{fit} \right)^2 / \sigma_{obs}^2}{N - f}, 
 \end{equation}
 \noindent
 with $\sigma_{obs}$ denoting the uncertainties in the observed velocity curve, $N$ the number of data points and \textcolor{black}{$N-f$} the degrees of freedom. In our case, $f=4$, as the two ends of the data set are used as fixed boundary points and two parameters are fitted. Due to our simpler, spherical model, our mass-luminosity ratio is not strictly physical (as a cylindrical model is usually preferred), but gives a plausible conversion factor for the fit. }
 \newpage
  \begin{figure}[ht]
		\centering
		\begin{floatrow}
			\ffigbox{%
				\includegraphics[height = 190pt]{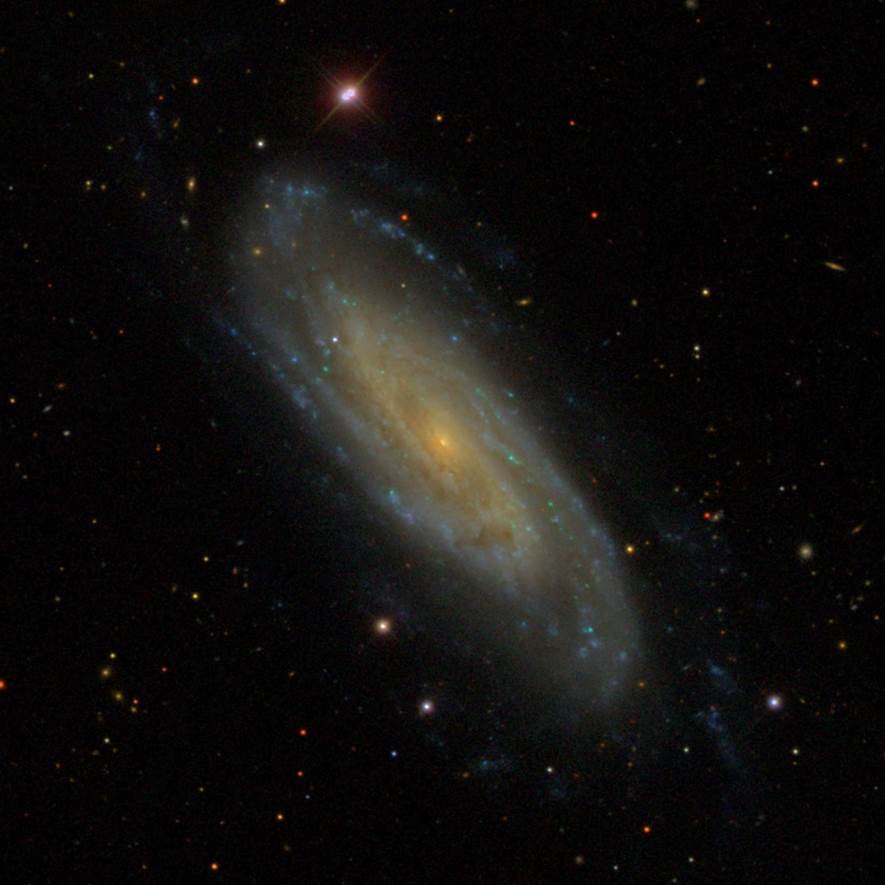}}{%
				\caption{NGC 3198 galaxy, source: Sloan Digital Sky  Survey.}%
			}
			\capbtabbox{%
				\bgroup
				\def\arraystretch{1.1}%
				\begin{tabular}{|c|c|}
					\hline
					\multicolumn{2}{|c|}{\textbf{ {\large NGC 3198}}}               \\ \hline
					\multicolumn{2}{|c|}{\textbf{\textcolor{black}{DC14 DM model}}}        \\ \hline
					\multicolumn{1}{|c|}{\textcolor{black}{$\log(\rho_s)$ [$\rho_s$: $ M_{\odot} pc^{-3}$]}}   &\textcolor{black}{ $-1.46 \pm 0.21$}\\ \hline 
 \textcolor{black}{$D$ [Mpc]}&\textcolor{black}{$10.47 \pm 1.17$}\\ \hline 
 \textcolor{black}{$r_s$ [$kpc$]}&\textcolor{black}{$5.77 \pm 1.23$}\\ \hline 
 \textcolor{black}{$i$ [deg]}&\textcolor{black}{$71.5 \pm 3.0$}\\ \hline
     				\multicolumn{1}{|c|}{\textcolor{black}{$M_*/L$ }   }    & \textcolor{black}{$0.40 \pm 0.09$}\\ \hline
                    \multicolumn{1}{|c|}{\textcolor{black}{$\chi^2_{\nu}$ }}    & \textcolor{black}{$1.26$}\\ \hline
					\multicolumn{2}{|c|}{\textcolor{black}{\textbf{MOND EFE}}} \\ \hline
					\multicolumn{1}{|c|}{\textcolor{black}{$\tilde{e}$}}     & \textcolor{black}{$0.055^{+0.012}_{-0.011}$}\\ \hline 
 \textcolor{black}{$D$ [Mpc]}&\textcolor{black}{$15.27^{+1.17}_{-1.08}$}\\ \hline 
 \textcolor{black}{$i$ [deg]}&\textcolor{black}{$75.62^{+2.73}_{-2.71}$}\\ \hline 
 \textcolor{black}{$M_{ \mathrm{gas}}/M_{\mathrm{HI}}$ }&\textcolor{black}{$1.36^{+0.13}_{-0.12}$}\\ \hline
     \multicolumn{1}{|c|}{\textcolor{black}{$M_{*, \mathrm{disk}}/L$ }}    & 
        \textcolor{black}{$0.43^{+0.04}_{-0.04}$}\\ \hline
    \multicolumn{1}{|c|}{\textcolor{black}{$\chi^2_{\nu}$ } }   & 
        \textcolor{black}{$1.62$}           \\ \hline
					\multicolumn{2}{|c|}{\textbf{Thermodynamic gravity}} \\  \hline
					\multicolumn{1}{|c|}{\textcolor{black}{$K$ [$10^{-5} s^2 km^{-2}$]}} & \textcolor{black}{$3.40 \pm 0.04$}        \\ \hline
     \multicolumn{1}{|c|}{\textcolor{black}{$M_*/L$ }   }    & \textcolor{black}{$0.762 \pm 0.007$}   \\ \hline
                        \multicolumn{1}{|c|}{\textcolor{black}{$\chi^2_{\nu}$ }   }   & \textcolor{black}{$1.29$}\\ \hline
				\end{tabular}
				\egroup
			}{%
				\caption{TG, \textcolor{black}{MOND EFE} and \textcolor{black}{DC14} DM parameters for galaxy NGC 3198 \textcolor{black}{\cite{Chae_2021, LiEta19a, Li_2020}} .}%
			}
		\end{floatrow}
	\end{figure}

 %\newpage
 
	\begin{figure}[H]
		\centering   
		\includegraphics[width= 0.8\textwidth]{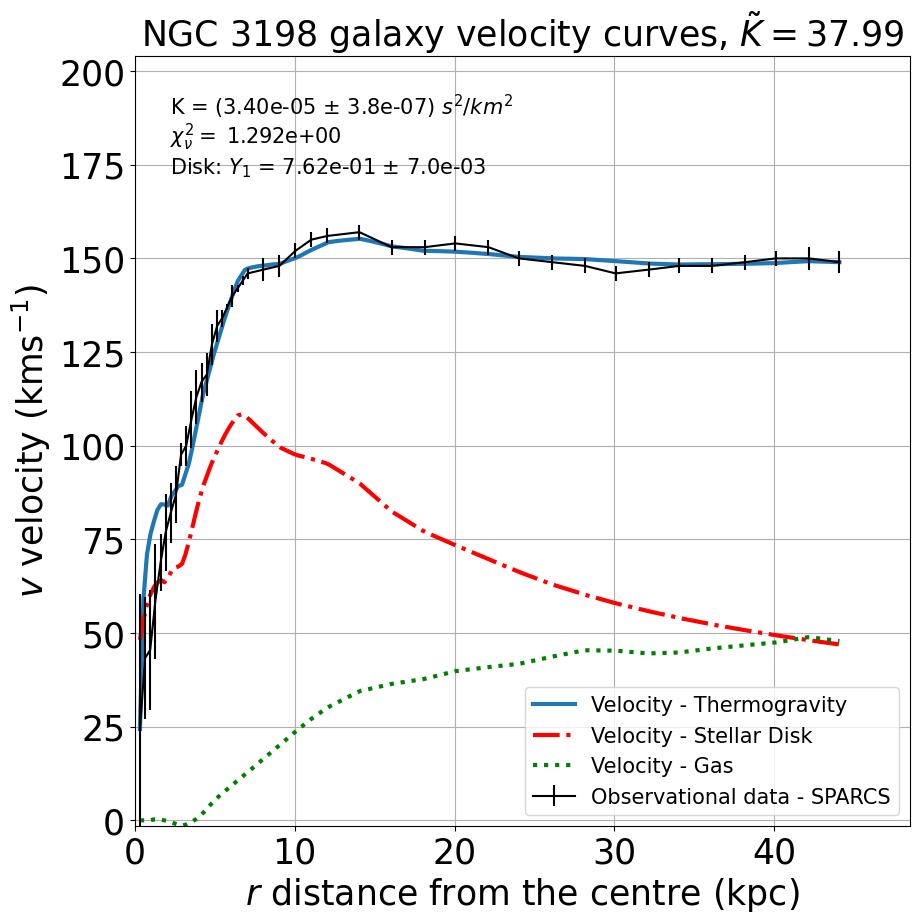}
  \includegraphics[width= 0.4\textwidth]{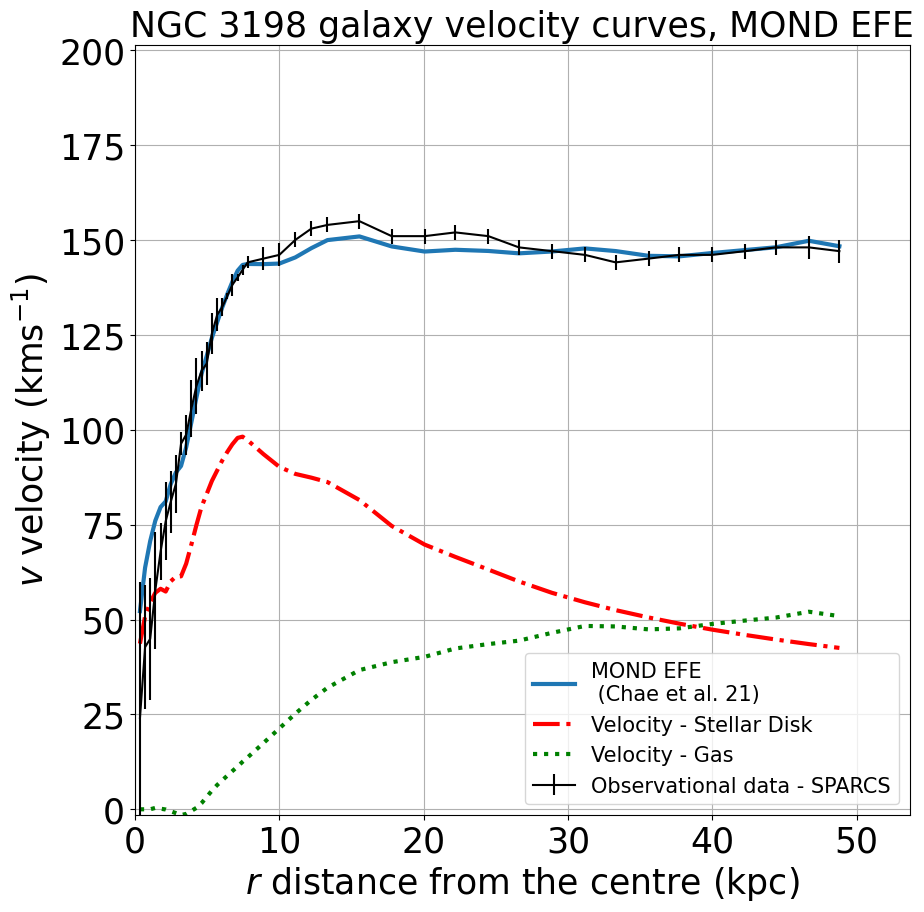}
  \includegraphics[width= 0.4\textwidth]{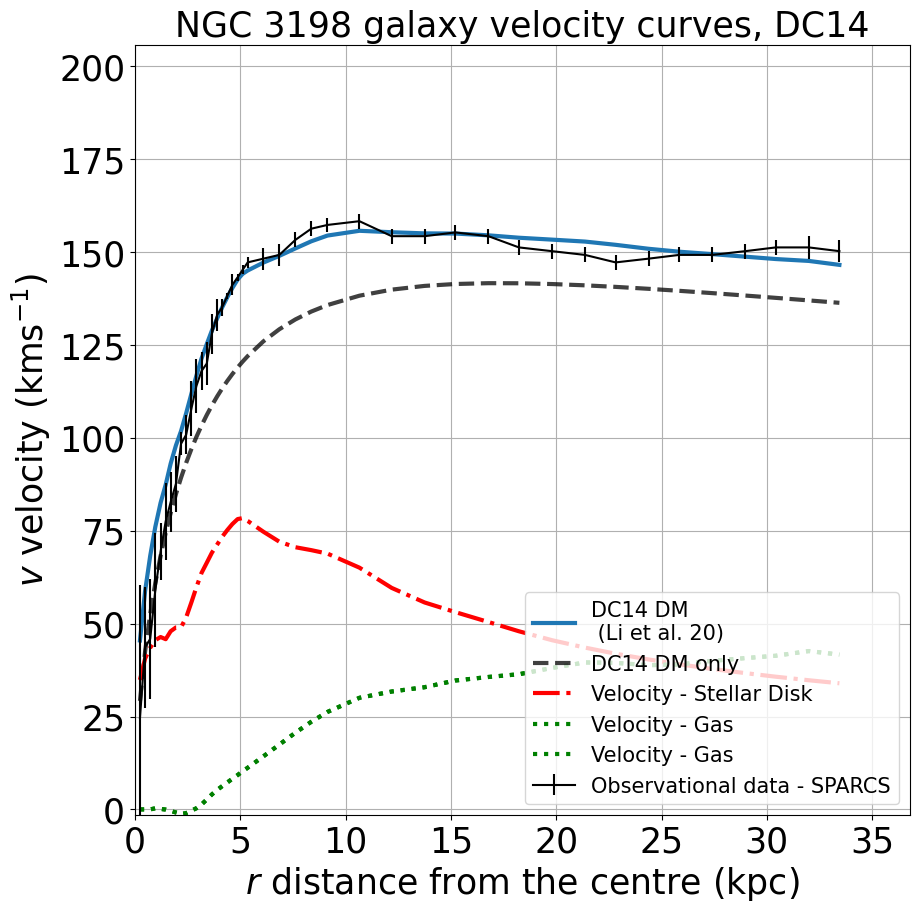} 
		\caption{Velocity curves from thermodynamic gravity, \textcolor{black}{MOND EFE from Chae et al. \cite{Chae_2021}} and \textcolor{black}{DC14 (flat prior) from \cite{LiEta19a, Li_2020}} dark matter for the NGC 3198 galaxy.}
		\label{fig:ngc3198-gal}
	\end{figure}

	\subsection{Discussion}
	
	NGC 3198 is a barred spiral galaxy known for its prototypical flat rotation curve, \cite{vanAlbada1985,KarEta15a,GenEta13a}. \textcolor{black}{It was chosen because all three investigated theories provide a good fit for the observed velocity data.} 
\textcolor{black}{It is remarkable that first MOND based rotation curve fits give acceptable result only with varying MOND acceleration parameters \cite{MONDvsDMhaloes,Wang_2021}. However, a more sophisticated modelling methodology considers the observation uncertainties and results in a good modelling also with the fixed, universal $a_0$, as in the MOND EFE model of Li et al. \cite{LiEta19a,Li_2020}. In our work, 
  thermodynamic gravity can very similarly match the rotation curves as the DC14 model DM and also as MOND EFE.}
 
 \textcolor{black}{In TG, the $K$ parameter arising from the coupling can naturally vary between galaxies, allowing for a more versatile approach, while still having a strong foundation.}

\subsection{Mass estimation}

 \textcolor{black}{The DC14 model total halo mass can be estimated as: 
 \begin{equation}
     M(r) = 4\pi r_s^3 \rho_s \frac{1}{\alpha} \left[ B(a, b+1, \epsilon) +B(a+1,b, \epsilon)  \right].
 \end{equation}}
 \noindent
 \textcolor{black}{ For the purposes of this estimation, the edge of the galaxy will be considered to be at the last observational velocity curve datapoint ($r_d$).} \textcolor{black}{In the case of $M_{DM}$, the galactic-distance-corrected distance is used.}  \textcolor{black}{Using the insights from eqs. \ref{g-in-vacuum} and \ref{C-to-M}, by considering the fitted K parameter for the galaxy, it is possible to estimate the apparent asymptotic mass of the galaxy by setting the acceleration inferred from the edge of the galaxy to equal to the vacuum solution:
	\begin{gather}
		\frac{v_{observed}(r_d)^2}{r_d} = \frac{1}{K r_d + C r_d^2}, \\
		C = \frac{1}{r_d} \left( \frac{1}{v_{observed}(r_d)^2} -K \right), \\
		M_{aa} =  \frac{v_{observed}(r_d)^2}{1-K v_{observed}(r_d)^2} \frac{r_d}{G} \label{apparent mass}.
	\end{gather}
}
	
 \textcolor{black}{In the case of $K = 0$, eq. \ref{apparent mass} should approximately yield the gravitating mass required for the observed velocity for the galaxy and is expected from the total mass inside $r_d$ with DM included. It can also be used to estimate the mass of the stellar disk (SD) and gas (HI) in the case of spherical symmetry, by using their respective velocities in the equation (and factoring in our fitted mass-luminosity ratio). Therefore, we can call $M_{aa, K=0}$ as {\em apparent total asymptotic mass} of the galaxy.
	The results are shown in Table \ref{mass-estimates-table}:}
	
	\begin{table}[h]
		\centering
		\bgroup
		\def\arraystretch{1.5}%
		\begin{tabular}{|c|c|c|c|c|c|c|c|}
			\hline
			\textbf{Galaxy} &
			\textbf{$M_{SD}$ } &
			\textbf{$M_{HI}$ } &
			\textcolor{black}{\textbf{$M_{BM}$ }} &
			\textcolor{black}{\textbf{$M_{DM}$ }} &
			\textbf{$M_{BM + DM}$ } &
			\textbf{$M_{aa, K=0}$ } &
			\textbf{$M_{aa}$ } \\ \hline
			NGC 3198 &
			\textcolor{black}{$34.1 $} &
			\textcolor{black}{$23.4 $} &
			\textcolor{black}{$57.5 $} &
			\textcolor{black}{$147.5$}& 
			\textcolor{black}{$205.0$}& 
			\textcolor{black}{$227.5 $} &
			\textcolor{black}{$927.8 $} \\ \hline
			\end{tabular}
		\egroup
		\caption{The mass components and estimates for \textcolor{black}{DC14} DM mass, all values are in the units of $10^{9} M_{\odot}$. $M_{SD}$ is the mass of the stellar components, $M_{HI}$ is the mass of neutral hydrogen, \textcolor{black}{$M_{BM}$} is the total mass of $M_{SD}$ and $M_{HI}$, $M_{DM}$ is the mass of the \textcolor{black}{DC14} model DM halo up to $r_d$, $M_{BM + DM}$ is the total mass, including baryonic and DM components, and $M_{aa}$ is the apparent asymptotic mass from the thermodynamic theory.}
		\label{mass-estimates-table}
	\end{table}
	
	We can see, that the total baryonic and dark matter mass of NGC 3198, $M_{BM+DM}$, is smaller than the apparent total mass of TG, but the apparent asymptotic mass with the galactic $K$ value, $M_{aa}$, while differing from the \textcolor{black}{DC14} DM mass, up to $r_d$ results in the same gravitational effect. The characteristic $R$ spatial distance of the crossover, as pictured on Fig. \ref{fig:double-crossover} and detailed in \cite{AbeVan22a} can be obtained as  
 \begin{equation}
     R = K GM_{aa, K=0}, 
 \end{equation}
 \noindent according to eq. \ref{reparam:g-in-vacuum}.

Using the $M_{aa, K=0} = 227.5 \cdot 10^9 M_{\odot}$ mass for this scale, the expected characteristic crossover distance of this galaxy is  $R \approx 33.3 $ kpc.
That is a crude estimate for distance where the flat velocity curve is expected to start declining and tending to a Newtonian one, which is determined by the vacuum solution of the Poisson equation with the apparent total asymptotic mass.
}
\subsection{Conclusions}
	
In this work, the analytical solutions of the nonlinear stationary field equation of TG were presented for the gravitational field in the cases of vacuum and constant matter density. Using the staggered grid-type discretisation, a numerical relaxation method was developed to solve the dissipative field equation \textcolor{black}{for arbitrary spherically symmetric mass distribution}. The stationary solution was obtained in the limit of the relaxation. The rotational velocity curve was calculated for galaxy NGC 3198 from the \textcolor{black}{SPARC} dataset, with only the baryonic matter contributing to the matter density source term {\cite{SPARC_master}. The $K$ parameter in the theory investigated by this work arises from the coupling between the gravitational and mechanical thermodynamic forces and fluxes and, thus, naturally varies across the galaxies, depending on their smaller-scale dynamics and characteristics.
	
Based on the developed numerical method, a next step is to extend the analysis to other galaxies of the \textcolor{black}{SPARC sample}, with a detailed comparison of the alternative models\textcolor{black}{, in particular for Fuzzy DM hydro solutions, similarly to \cite{Chae_2022} and \cite{SPARC_DM_over_MOND}}. It is also natural to investigate thermodynamic gravity on the scale of galaxy clusters. It may be the case that on much larger scales, the theory has effects similar to Newtonian or general relativistic gravity, except the apparent mass is larger due to the thermodynamic coupling, with a crossover between the regimes \cite{AbeVan22a}. 

\textcolor{black}{The investigation of other challenging observed effects, such as gravitational lensing and the description of the fluid mechanics of the intracluster gas in the Bullet Cluster (1E 0657-56) are necessary to estimate the feasibility of a novel theory of gravity. In this respect, TG seems to be promising because the nonlinear field equation and the strong hydrodynamic coupling indicate that the lensing convergence can be non-zero where there is no projected matter, contrary to Einsteinian gravity, \cite{AngEta06a}. Also, the success of self-interacting dark matter investigations indicate the modelling potential of hydrodynamic approaches, \cite{BulletClusterSIDM}. Naturally, the Bullet Cluster, together with several other cosmological observations that have already been studied in the context of DM and MOND, require further research. The constitutive character of TG is an advantage when explaining and substituting the rich phenomenology of DM with modified gravity, \cite{Salucci2019}.}

For simplicity and analytical comparisons, the numerical method worked with a spherically symmetric model of mass distribution, although the baryonic mass of the non-elliptical galaxies usually \textcolor{black}{is described by exponential disks with different scale heights in the axial $z$ and radial directions.}
\textcolor{black}{However, the solution of nonlinear field equations for nonspherical mass distributions can be mathemathically demanding \cite{ChaeMilgrom_2022} and certain physical considerations should be taken into account \cite{GIORDANO201962, chavanis2021selfgravitating, giordano2023fluid}}.

	\vspace*{\fill}
	
	\begin{center}
		\section*{Acknowledgements}
	\end{center}
	
	We would like to thank Professor W. J. G. de Blok for providing us with the THINGS  data \cite{de_Blok_2008}.  We would also like to thank S. Abe and \textcolor{black}{ 
R. Trasarti-Battistoni} for providing insightful remarks regarding the theory. \textcolor{black}{We would like to thank the anonymous referees for their constructive and helpful feedback.}

 The work was supported by the grants National Research, Development and Innovation Office –  FK134277. The research reported in this paper is part of project no. BME-NVA-02, implemented with the support provided by the Ministry of Innovation and Technology of Hungary from the National Research, Development and Innovation Fund, financed under the TKP2021 funding scheme.
 
\section*{Data Availability}

The data for the galaxy NGC 3198 used in this paper is based on the \textcolor{black}{SPARC sample, \cite{SPARC_master}.}
	
%	\newpage
	\bibliographystyle{unsrt}
%	\bibliography{references}

\begin{thebibliography}{10}	
	\bibitem{Bertone_2018}
	G.~Bertone and T.~M.~P. Tait.
	\newblock A new era in the search for dark matter.
	\newblock {\em Nature}, 562(7725):51--56, oct 2018.
	
	\bibitem{x17}
	A.~J. Krasznahorkay, M.~Csatl\'os, L.~Csige, J.~Guly\'as, A.~Krasznahorkay,
	B.~M. Nyak\'o, I.~Rajta, J.~Tim\'ar, I.~Vajda, and N.~J. Sas.
	\newblock New anomaly observed in $^{4}\mathrm{He}$ supports the existence of
	the hypothetical {X}17 particle.
	\newblock {\em Phys. Rev. C}, 104:044003, Oct 2021.
	
	\bibitem{neutrinomass}
	A.~Datta, R.~Roshan, and A.~Sil.
	\newblock Imprint of the {S}eesaw {M}echanism on {F}eebly {I}nteracting {D}ark
	{M}atter and the {B}aryon {A}symmetry.
	\newblock {\em Phys. Rev. Lett.}, 127:231801, Dec 2021.
	
	\bibitem{Famaey2012}
	B.~Famaey and S.~S. McGaugh.
	\newblock Modified {N}ewtonian {D}ynamics ({MOND}): {O}bservational
	{P}henomenology and {R}elativistic {E}xtensions.
	\newblock {\em Living Reviews in Relativity}, 15(1), 2012.
	
	\bibitem{KatzEtA17}
	Harley Katz, Federico Lelli, Stacy~S. McGaugh, Arianna Di~Cintio, Chris~B.
	Brook, and James~M. Schombert.
	\newblock {Testing feedback-modified dark matter haloes with galaxy rotation
		curves: estimation of halo parameters and consistency with $\Lambda$CDM
		scaling relations}.
	\newblock {\em Monthly Notices of the Royal Astronomical Society},
	466(2):1648--1668, 12 2016.
	
	\bibitem{SPARC_DM_over_MOND}
	Mariia Khelashvili, Anton Rudakovskyi, and Sabine Hossenfelder.
	\newblock {SPARC} galaxies prefer {D}ark {M}atter over {MOND}, 2024.
	
	\bibitem{Lelli_2017}
	F.~Lelli, S.~S. McGaugh, J.~M. Schombert, and M.~S. Pawlowski.
	\newblock One {L}aw to {R}ule {T}hem {A}ll: {T}he {R}adial {A}cceleration
	{R}elation of {G}alaxies.
	\newblock {\em The Astrophysical Journal}, 836(2):152, feb 2017.
	
	\bibitem{Li2018}
	P.~Li, F.~Lelli, S.~McGaugh, and J.~Schombert.
	\newblock Fitting the radial acceleration relation to individual {SPARC}
	galaxies.
	\newblock {\em Astronomy {\&} Astrophysics}, 615:A3, 2018.
	
	\bibitem{lambdacdm-challenges}
	James~S. Bullock and Michael Boylan-Kolchin.
	\newblock Small-{S}cale {C}hallenges to the {$\Lambda$}{C}{D}{M} {P}aradigm.
	\newblock {\em Annual Review of Astronomy and Astrophysics}, 55(1):343--387,
	2017.
	
	\bibitem{Bol21a}
	P.~Boldrini.
	\newblock The cusp--core problem in gas-poor dwarf spheroidal galaxies.
	\newblock {\em Galaxies}, 10(01):5, 2021.
	
	\bibitem{BerKho15a}
	L.~Berezhiani and J.~Khoury.
	\newblock Theory of dark matter superfluidity.
	\newblock {\em Physical Review D}, 92(10):103510, 2015.
	
	\bibitem{covariant-emergent-grav}
	S.~Hossenfelder.
	\newblock Covariant version of {V}erlinde's emergent gravity.
	\newblock {\em Phys. Rev. D}, 95:124018, Jun 2017.
	
	\bibitem{Zeilinger23}
	Horst Foidl, Tanja Rindler-Daller, and Werner~W. Zeilinger.
	\newblock Halo formation and evolution in scalar field dark matter and cold
	dark matter: {N}ew insights from the fluid approach.
	\newblock {\em Phys. Rev. D}, 108:043012, Aug 2023.
	
	\bibitem{Mina22}
	{Mina, Mattia}, {Mota, David F.}, and {Winther, Hans A.}
	\newblock Solitons in the dark: First approach to non-linear structure
	formation with fuzzy dark matter.
	\newblock {\em A\&A}, 662:A29, 2022.
	
	\bibitem{Zlo22a}
	K.G. Zloshchastiev.
	\newblock Galaxy rotation curves in superfluid vacuum theory.
	\newblock {\em Pramana}, 97(1):2, 2022.
	
	\bibitem{Sco22a}
	T.~C. Scott.
	\newblock From {Modified Newtonian Dynamics} to superfluid vacuum theory.
	\newblock {\em Entropy}, 25(1):12, 2022.
	
	\bibitem{MaeGue20a1}
	A.~Maeder and V.~G. Gueorguiev.
	\newblock The scale-invariant vacuum {(SIV)} theory: A possible origin of dark
	matter and dark energy.
	\newblock {\em Universe}, 6(3):46, 2020.
	
	\bibitem{aqual}
	J.~{Bekenstein} and M.~{Milgrom}.
	\newblock {Does the missing mass problem signal the breakdown of {N}ewtonian
		gravity?}
	\newblock {\em The Astrophysical Journal}, 286:7--14, November 1984.
	
	\bibitem{refId0}
	P.~Kroupa, B.~Famaey, K.~S. de~Boer, J.~Dabringhausen, M.~S. Pawlowski, C.~M.
	Boily, H.~Jerjen, D.~Forbes, G.~Hensler, and M.~Metz.
	\newblock Local-{G}roup tests of dark-matter concordance cosmology - {T}owards
	a new paradigm for structure formation.
	\newblock {\em A\&A}, 523:A32, 2010.
	
	\bibitem{dodelson}
	S.~Dodelson.
	\newblock The real problem with {MOND}.
	\newblock {\em International Journal of Modern Physics D}, 20(14):2749--2753,
	2011.
	
	\bibitem{Ver17a}
	E.~P. Verlinde.
	\newblock {Emergent {G}ravity and the {D}ark {U}niverse}.
	\newblock {\em SciPost Phys.}, 2:016, 2017.
	
	\bibitem{Hos17a}
	S.~Hossenfelder.
	\newblock Covariant version of verlinde’s emergent gravity.
	\newblock {\em Physical Review D}, 95(12):124018, 2017.
	
	\bibitem{vanabe}
	P.~Ván and S.~Abe.
	\newblock Emergence of extended {N}ewtonian gravity from thermodynamics.
	\newblock {\em Physica A: Statistical Mechanics and its Applications},
	588:126505, 2022.
	
	\bibitem{AbeVan22a}
	S.~Abe and P.~Ván.
	\newblock Crossover in {E}xtended {N}ewtonian {G}ravity {E}merging from
	{T}hermodynamics.
	\newblock {\em Symmetry}, 14(5), 2022.
	
	\bibitem{Van23a}
	P.~V\'an.
	\newblock Holographic fluids: a thermodynamic road to quantum physics.
	\newblock {\em Physics of Fluids}, 35(5):057105, 2023.
	\newblock arXiv:2301.07177v2.
	
	\bibitem{Logotropic_Chavanis22}
	Pierre-Henri Chavanis.
	\newblock New logotropic model based on a complex scalar field with a
	logarithmic potential.
	\newblock {\em Phys. Rev. D}, 106:063525, Sep 2022.
	
	\bibitem{Giu97a}
	D.~Giulini.
	\newblock Consistently implementing the field self-energy in {N}ewtonian
	gravity.
	\newblock {\em Physics Letters A}, 232(3-4):165--170, 1997.
	
	\bibitem{Sivaram2020}
	C.~Sivaram, A.~Kenath, and L.~Rebecca.
	\newblock {MOND}, {MONG}, {MORG} as alternatives to dark matter and dark
	energy, and consequences for cosmic structures.
	\newblock {\em Journal of Astrophysics and Astronomy}, 41(1), February 2020.
	
	\bibitem{rebecca23}
	L.~Rebecca, A.~Kenath, and C.~Sivaram.
	\newblock Baryonic matter abundance in the framework of mong.
	\newblock {\em Physical Sciences Forum}, 7(1), 2023.
	
	\bibitem{DIOSI20131782}
	L.~Diósi.
	\newblock Note on possible emergence time of {N}ewtonian gravity.
	\newblock {\em Physics Letters A}, 377(31):1782 -- 1783, 2013.
	
	\bibitem{nummodszerek}
	Kovács~R. és Józsa~V.
	\newblock {\em Bevezetés a numerikus módszerekbe}.
	\newblock Akadémiai Kiadó, 2019.
	
	\bibitem{RieEta18a}
	{\'A}.~Rieth, R.~Kov{\'a}cs, and T.~F{\"u}l{\"o}p.
	\newblock Implicit numerical schemes for generalized heat conduction equations.
	\newblock {\em International Journal of Heat and Mass Transfer},
	126:1177--1182, 2018.
	
	\bibitem{PozsEta20a}
	{\'A}.~Pozs{\'a}r, M.~Sz{\"u}cs, R.~Kov{\'a}cs, and T.~F{\"u}l{\"o}p.
	\newblock Four spacetime dimensional simulation of rheological waves in solids
	and the merits of thermodynamics.
	\newblock {\em Entropy}, 22(12):1376, 2020.
	
	\bibitem{SPARC_master}
	Federico Lelli, Stacy~S. McGaugh, and James~M. Schombert.
	\newblock Sparc: Mass models for 175 disk galaxies with spitzer photometry and
	accurate rotation curves.
	\newblock {\em The Astronomical Journal}, 152(6):157, nov 2016.
	
	\bibitem{LiEta19a}
	Pengfei Li, F.~Lelli, S.~S. McGaugh, N.~Starkman, and J.~M. Schombert.
	\newblock A constant characteristic volume density of dark matter haloes from
	{SPARC} rotation curve fits.
	\newblock {\em Monthly Notices of the Royal Astronomical Society},
	482(4):5106--5124, 2019.
	
	\bibitem{Li_2020}
	Pengfei Li, Federico Lelli, Stacy McGaugh, and James Schombert.
	\newblock A {C}omprehensive {C}atalog of {D}ark {M}atter {H}alo {M}odels for
	{SPARC} {G}alaxies.
	\newblock {\em The Astrophysical Journal Supplement Series}, 247(1):31, mar
	2020.
	
	\bibitem{Chae_2021}
	Kyu-Hyun Chae, Harry Desmond, Federico Lelli, Stacy~S. McGaugh, and James~M.
	Schombert.
	\newblock Testing the {S}trong {E}quivalence {P}rinciple. {II}. {R}elating the
	{E}xternal {F}ield {E}ffect in {G}alaxy {R}otation {C}urves to the
	{L}arge-scale {S}tructure of the {U}niverse.
	\newblock {\em The Astrophysical Journal}, 921(2):104, nov 2021.
	
	\bibitem{Zhao_2006}
	H.~S. Zhao and B.~Famaey.
	\newblock Refining the {MOND} {I}nterpolating {F}unction and {TeVeS}
	{L}agrangian.
	\newblock {\em The Astrophysical Journal}, 638(1):L9, jan 2006.
	
	\bibitem{Wang_2021}
	Lin Wang and Da-Ming Chen.
	\newblock Comparison of modeling {SPARC} spiral galaxies’ rotation curves:
	halo models vs. {MOND}.
	\newblock {\em Research in Astronomy and Astrophysics}, 21(11):271, dec 2021.
	
	\bibitem{vanAlbada1985}
	T.~S. van Albada, J.~N. Bahcall, K.~Begeman, and R.~Sancisi.
	\newblock Distribution of dark matter in the spiral galaxy {NGC} 3198.
	\newblock {\em The Astrophysical Journal}, 295:305, August 1985.
	
	\bibitem{KarEta15a}
	E.V. Karukes, P.~Salucci, and G.~Gentile.
	\newblock The dark matter distribution in the spiral {NGC} 3198 out to {0.22
		Rvir}.
	\newblock {\em Astronomy \& Astrophysics}, 578:A13, 2015.
	
	\bibitem{GenEta13a}
	G.~Gentile, G.I.G. J{\'o}zsa, P.~Serra, G.H. Heald, W.J.G. de~Blok,
	F.~Fraternali, M.T. Patterson, R.A.M. Walterbos, and T.~Oosterloo.
	\newblock {HALOGAS: Extraplanar gas in NGC 3198}.
	\newblock {\em Astronomy \& Astrophysics}, 554:A125, 2013.
	
	\bibitem{MONDvsDMhaloes}
	T.~H. Randriamampandry and C.~Carignan.
	\newblock {Galaxy mass models: MOND versus dark matter haloes}.
	\newblock {\em Monthly Notices of the Royal Astronomical Society},
	439(2):2132--2145, 2014.
	
	\bibitem{Chae_2022}
	Kyu-Hyun Chae.
	\newblock Distinguishing dark matter, modified gravity, and modified inertia
	with the inner and outer parts of galactic rotation curves.
	\newblock {\em The Astrophysical Journal}, 941(1):55, dec 2022.
	
	\bibitem{AngEta06a}
	G.~W. Angus, B.~Famaey, and HongSheng Zhao.
	\newblock Can {MOND take a bullet? Analytical comparisons of three versions of
		MOND} beyond spherical symmetry.
	\newblock {\em Monthly Notices of the Royal Astronomical Society},
	371(1):138--146, 2006.
	
	\bibitem{BulletClusterSIDM}
	Andrew Robertson, Richard Massey, and Vincent Eke.
	\newblock {What does the {B}ullet {C}luster tell us about self-interacting dark
		matter?}
	\newblock {\em Monthly Notices of the Royal Astronomical Society},
	465(1):569--587, 10 2016.
	
	\bibitem{Salucci2019}
	P.~Salucci.
	\newblock The distribution of dark matter in galaxies.
	\newblock {\em The Astronomy and Astrophysics Review}, 27(1), February 2019.
	
	\bibitem{ChaeMilgrom_2022}
	Kyu-Hyun Chae and Mordehai Milgrom.
	\newblock Numerical {S}olutions of the {E}xternal {F}ield {E}ffect on the
	{R}adial {A}cceleration in {D}isk {G}alaxies.
	\newblock {\em The Astrophysical Journal}, 928(1):24, mar 2022.
	
	\bibitem{GIORDANO201962}
	D.~Giordano, P.~Amodio, F.~Iavernaro, A.~Labianca, M.~Lazzo, F.~Mazzia, and
	L.~Pisani.
	\newblock Fluid statics of a self-gravitating perfect-gas isothermal sphere.
	\newblock {\em European Journal of Mechanics - B/Fluids}, 78:62--87, 2019.
	
	\bibitem{chavanis2021selfgravitating}
	P.-H. Chavanis.
	\newblock The self-gravitating {F}ermi gas in {N}ewtonian gravity and general
	relativity, 2021.
	
	\bibitem{giordano2023fluid}
	Domenico Giordano, Pierluigi Amodio, Felice Iavernaro, Francesca Mazzia, Péter
	Ván, and Mátyás Szücs.
	\newblock Fluid statics of a self-gravitating isothermal sphere of van der
	{W}aals' gas, 2024.
	\newblock arXiv:2311.12150, to be published in Physics of Fluids.
	
	\bibitem{de_Blok_2008}
	W.~J.~G. de~Blok, F.~Walter, E.~Brinks, C.~Trachternach, S-H. Oh, and R.~C.
	Kennicutt.
	\newblock High-{R}esolution {R}otation {C}urves and {G}alaxy {M}ass {M}odels
	from {T}{H}{I}{N}{G}{S}.
	\newblock {\em The Astronomical Journal}, 136(6):2648--2719, 2008.
	
\end{thebibliography}

\end{document}